\documentstyle[graphics]{mn}
\begin{document}
\title[Sweeping and Annealing of Disc Galaxies]
{Multi-Stage Three Dimensional Sweeping and Annealing of Disc Galaxies
in Clusters}
\author[S. Schulz and C. Struck]
{Steven Schulz and Curtis Struck \\
Dept. of Physics and Astronomy, Iowa State University,
Ames, IA 50011  USA}
\maketitle
\begin{abstract}

We present new three dimensional, hydrodynamic simulations of the ram
pressure stripping of disc galaxies via interaction with an hot
intracluster medium (ICM). The simulations were carried with the
smoothed-particle hydrodynamics, adaptive mesh 'Hydra' code
(SPH-A$P^3$M), with model galaxies consisting of dark halo, and gas
and stellar disc components.  The simulations also include radiative
cooling, which is important for keeping the warm, diffuse gas of
moderate density from being unrealistically heated by the ICM.  We
examine the role that wind velocity, density and galaxy tilt play in
gas stripping.  We include cases with lower ram pressures than other recent
studies.

In accord with previous studies we find that low column density gas is
promptly removed from the outer disc.  However, we also find that not
all of the gas stripped from the disc escapes immediately from the
halo, some of material can linger for times of order $10^8$ yr.  We
use a simple analytic model to demonstrate that gas elements in the ICM
wind feel an effective potential with a minimum displaced downstream
from the halo centre.

The onset of the ICM wind has a profound effect on the disc gas that
is not immediately stripped.  This remnant disc is displaced relative
to the halo center and compressed.  This can trigger gravitational
instability and the formation of numerous flocculent spirals.  These
waves transport angular momentum outward, resulting in further
compression of the inner disc and the formation of a prominent gas
ring.  This 'annealing' process makes the inner disc, which contains
much of the total gas mass, resistant to further stripping, but
presumably susceptible to global starbursts.  Spirals in the outer
disc stretch, shear and are eventually stripped on timescales of a few
times $10^8$ yr., after which time mass and angular momentum loss
effectly cease.

For inclined galaxies, these effects are considerably modified over
the same timescale.  The amount of mass loss is reduced.  In addition,
we find that a higher galaxy tilt couples the wind and the rotating
disc, and produces a higher degree of angular momentum removal.
Temperature and line of sight velocity maps from several of the
simulations are presented for comparison to observation.

When the mass loss and annealing processes go to completion we find that
the total amount of mass lost from a fixed target galaxy is well fit
by a simple power-law function of a dimensionless parameter that
combines the ram pressure and internal properties of the galaxy.
Ramifications for the cluster galaxy evolution are discussed.
\end{abstract}


\section{INTRODUCTION: HIGHLIGHTS OF RAM PRESSURE STRIPPING THEORY}

In this paper we present new three dimensional numerical models of
stripping from gas-rich disc galaxies in clusters, with radiative
cooling included.  We begin with a brief review the literature to put
this work in context.

Gunn and Gott (1972) initiated the study of ram pressure stripping of
disc gas in cluster galaxies by the intracluster medium (ICM) in their
seminal study of the formation and evolution of galaxy clusters.
Specifically, they estimated how the ICM ram pressure compared to the
local gravitational binding of the gas at a representative point in a
typical galaxy disc (e.g., the solar radius in the Milky Way), and
found that stripping was likely to be very effective, and by
implication prompt.  The Gunn and Gott gravitational binding estimate
included only the stellar and gas disc components, but no halo.

The first numerical hydrodynamical simulations were published within a
few years by Lea and De Young (1976), and Gisler (1976).  Lea and De
Young found that 80-90\% of the gas in the model galaxy was stripped
within one crossing of a cluster like Coma, or within about $10^9$
yrs.  However, calculations were two-dimensional, used a beam scheme
algorithm, and an adiabatic equation of state.  The gravity of the
early type model galaxy was assumed to be dominated by a spherical,
isotropic stellar potential, with only a single hot gas component,
initially in hydrostatic equilibrium.  Gisler also considered the gas
mass evolution in cluster ellipticals, including the effects of: gas
input by stellar mass loss, gas loss in galactic winds, and ICM
stripping.  His two-dimensional, fluid-in-cell simulations also
produced substantial stripping.

In the late 1970s and early 1980s, satellite X-ray observations of the
hot gas in clusters, and of galaxies within the nearest clusters were
published (see review of Forman and Jones 1982).  The X-ray emission
characteristics were commonly interpreted in terms of stripping
models, and provided support for those models as well as a more
precise picture of the ICM. 

At the same time, evidence mounted that all types of spirals in
clusters are gas poor relative to their neighbors in the field,
specifically that they are HI deficient (e.g., Bothun 1982, Giovanelli
and Haynes 1983).  Systematic dependences on galaxy position and
velocity relative to cluster core values supported stripping as the
cause of the deficits, rather than formation processes. More recent
work has confirmed, and greatly extended, these results, e.g., in the
Virgo cluster (Cayatte et al. 1990), and Coma (Bravo-Alfaro et
al. 2000, also see Van Gorkom's 1996 review of HI in cluster
galaxies).

Kenney and Young (1986, 1988) noted that molecular gas, typically
found at smaller radii in galaxy discs than the HI, is not stripped.
This showed that stripping is not strong enough to remove the denser
gas in the inner discs of spirals, and is not as complete as we might
have expected from the earliest models.

Farouki and Shapiro (1980) carried out three-dimensional particle
simulations to study the thickening of stellar discs following
stripping.  While their primary goal was to investigate the SO origin
question, and their models had only 100 ``gas'' particles, their
simulations were the first on stripping from discs. They found: that
only the outer parts of discs would be swept in some cases, that the
amount of stripping depended on ICM parameters, and that stripping
might not always occur promptly.  Their statement that - ``The action
of a strong restoring force on gas clouds venturing from the disc
following ejection gives rise to complicated motions in the
simulations'' - is prescient and relevant to results we present below.

The simulations and analytic work of Takeda, Nulsen \& Fabian (1984),
advanced the theory in several areas.  They also carried out
two-dimensional simulations of spherical galaxies with a spherical gas
distribution using a fluid-in-cell algorithm.  In their simulations,
they assumed the galaxy traveled on a nearly radial orbit, and used a
cluster model with a hydrostatic, isothermal gas distribution, to
derive density and velocity variations along the galaxy orbit.  At the
outermost excursion, the galaxy velocity and ICM density are low,
stripping is unimportant, and the galaxy rebuilds its gas supply via
stellar mass loss.

Takeda et al. identified three stripping regimes: 1) prompt, in which
ram pressure immediately overwhelms gravitational binding and removes
the gas, 2) continuous, in which it takes the ICM wind some time to
provide the momentum to remove the gas (and the gas may be partially
replenished on a comparable timescale, or even accrete with low
relative velocities), 3) cyclical, in which gas is stripped in the
cluster core, but replenished in the outer cluster.

Nulsen (1982), in a review of both hydrodynamic and kinetic processes
in stripping concluded that turbulent shear viscosity would play an
important role.  However, Takeda et al.\ state that the shear
viscosity included in their models did not affect their basic results,
such as the existence of different stripping regimes.  Recently, Mori
\& Burkert (2000) have resolved Kelvin Helmholtz turbulence in models
of stripping from dwarf galaxies. While they find that it
significantly enhances mass loss from dwarfs, the characteristic
estimated timescale for the process is very long in normal, disc
galaxies.  

The work of Gaetz, Salpeter \& Shaviv (1987, also see Shaviv and
Salpeter, 1982) introduced more physical processes into models of
stripping from spherical galaxies.  These authors used a second order
Eulerian hydrodynamics code to carry out two-dimensional simulations,
with a steady-state cooling curve, and algorithms for star formation,
gas consumption and replenishment.  The various processes were
characterized by 5 timescales and their ratios.  They found that with
these competing processes included it was possible to achieve a steady
state, for galaxies moving through a uniform medium.  The degree of
stripping depended on the parameter
$n_{icm}v_{gal}^{2.4}/{\dot{M}}_{rep},$ where the variables are the
density of the ICM, the velocity of the galaxy through the medium, and
the constant replenishment rate, respectively.  Portnoy, Pistinner, \&
Shaviv (1993) updated and extended the models of Gaetz et al., and
provided fitting formulae for the effects of various parameters.

More recent high resolution models of stripping from spherical
galaxies have been presented by Balsara, Livio, \& O'Dea (1994), and
Stevens, Acreman, and Ponman (1999).  The models of both groups were
calculated with two dimensional, PPM codes, which included the effects
of star formation and replenishment.  Like Takeda et al., Stevens et
al. found a number of stripping modes, including a cyclic one.  Both
works had high spatial resolution in the wake, and they argued that
the numerical viscosity of previous works limited their ability to
resolve shocks.  However, some contradictory results, and questions
about kinetic effects and the need for a two-fluid prescription (see
Portnoy, et al. 193), leave the theory of stripping of hot gas from
spherical galaxies in a rather ambiguous state.

Kritsuk (1983) examined the stripping of different types of
interstellar clouds, located at different galactic radii, and with ICM
flows at a range of angles relative to the disc plane.  Kritsuk found
that diffuse HI clouds would be readily swept in most cases, while
dense molecular clouds would have a low stripping probability.  His
approach was to derive an equation of motion for clouds, subject to
gravity from the stellar disc and ram pressure forces, including mass
accretion and ablation.  The latter effects, and cloud structural
evolution were treated very approximately.

From the cloud equation of motion he derived an azimuthally dependent
stripping radius in the galaxy disc.  This work provided information
about three-dimensional stripping, and except for one run of Farouki
and Shapiro, is the first time the effects of inclination where
examined.  For example, Kritsuk's Figure 2 shows a strong coupling
between disc rotation and the ICM wind in highly inclined cases, a
result that will be emphasized in the new models below.

  A number of recent observations of asymmetric, extra-disc gas
distributions, or wakes, in cluster galaxies give a direct view of the
stripping process.  E.g., van Driel and van Woerden (1989) mapped the
HI emission from NGC 4694, a Virgo cluster lenticular, and located
most of the emission in an HI tail, 36 kpc. in length. They favor a
stripping explanation for the origin of the tail, though the presence
of a small dwarf galaxy makes an interaction origin possible.

  M96 (NGC 4406), a Virgo elliptical with X-ray, HI and infrared
emission from an asymmetric plume (White et al. 1991) has long been
considered a possible example of stripping.  The giant Virgo
elliptical NGC 4472 seems to be in a very similar circumstance.  X-ray
observations show an asymmetic gas plume of comparable size to the
optical galaxy, and an apparent bow shock on the opposite side (Irwin
and Sarazin 1996).

  These examples confirm the general morphology of stripping suggested
by the early hydrodynamic models.  The presence of gas within cluster
ellipticals also supports the conclusions that either gas is not
entirely stripped from early-type galaxies, or replenishment is
important, and perhaps leads to a steady or slowing changing
state. 

The discovery in the last decade that many clusters continue to
accrete groups of galaxies, or merge with other clusters, has changed
the conceptual framework of the field.  Continuing infall of spirals
into large clusters can account for diverse disc types in clusters,
and provides the opportunity for more precisely constraining the
stripping process.  For example, Phookun and Mundy (1995) presented a
well resolved HI map of the Virgo spiral NGC 4654, which showed a very
asymmetric HI distribution, with the NW side compacted, as expected
for a bow shock, and a long tail on the SE side. They argue that these
asymmetries are the result of the combined effects of ram pressure and
rotation if the cluster 'wind' is not face-on to the disc. Indeed, we
find a very similar morphology in inclinded disc models we present
below. Vollmer et al. (1999) find similar, asymmetries in the Virgo
disc galaxy NGC 4548.  Both objects have kinematic asymmetries as
well.

Kenney and Koopman (1999) also found morphological distortions in
their H$\alpha$ imagery of the Virgo spiral NGC 4522.  This is also
true of the HI and radio continuum emission, according to van Gorkom
(2000, private comm.)). Here the cluster wind (with a projected
velocity of 1300 km/s) seems to be nearly face-on to the disc, which
is observed nearly edge-on.  The SE side appears compressed, while
extraplanar filaments of gas emission are found on the NW side.  (Also
see Gavazzi et al. 1995, and Chaname\' , Infante, \& Reisenegger 2000,
for more objects with similar morphologies.)

Asymmetries are also common in the cluster galaxy HI surveys
cited above, and Rubin, Waterman, \& Kenney (1999) find, from optical
spectroscopy, that kinematic disturbances are common in Virgo disc
galaxies.

These observational results stimulated new (three dimensional)
modeling studies of sweeping from disc galaxies, including: tree-SPH
models of Abadi, Moore, \& Bower (1999), sticky particle models of
Vollmer (2000), and simulations made with a high resolution Eulerian
finite difference code by Quilis et al. (2000).  With an extensive,
pre-existing gas disc, the continuum, hydrodynamic limit should be a
good approximation for calculating the formation and structure of the
bow shock.  Abadi et al. and Quilis et al. both argue that stripping
from disc galaxies is very prompt. Abadi et al. find that up to 80\%
of the gas of a spiral galaxy can be lost in about $10^7$ yr., in the
core of a dense cluster like Coma.

Vollmer also found that stripping occurred in prompt events.
Vollmer's models were run with a time-varying wind to represent
passage through different parts of the cluster. They found that for
galaxies on radial orbits, a great deal of material could be pushed
out into long tails in the wake, but a significant amount of this
material falls back as the galaxy moved out of the cluster core
(especially for higher angles between the wind and the disc rotation
axis).  A corollary of this is that the details of stripping depend
strongly on the type of orbit a galaxy pursues through the cluster (as
Takeda et al. concluded for spherical galaxies).  Vollmer use a rather
ad hoc prescription for the wind pressure, which only approximates
pressure effects, so detailed results should be treated with caution.

Abadi et al. carried out a grid of runs in which constant ICM winds
impact multicomponent model galaxies (bulge/disc/halo) and compared
them to semi-analytic stripping criteria like those of Gunn and Gott, but
generalized to multicomponent galaxies.  Quilis et al. use a finite
difference code to achieve higher spatial resolution, but present only
a couple simulations of sweeping from galaxies with central gas holes.

Our simulations were generally run for a time of $6 \times 10^8$ yrs.,
comparable to those of Vollmer, but those of Abadi et al. were run for
only slightly more than $10^8$. Although they emphasize that most of
the stripping has already occurred by this time, their figures show
some disc gas remains nearby.  They also show slightly increasing gas
mass in the disc at the end of some runs.  Thus, some fallback or
removal must be completed on a longer timescale, as in Vollmer's and
our models.

  More generally, we find that the dynamics of stripped gas is quite
complex on timescales of a few times $10^8$ yrs. (like Farouki \&
Shapiro, and Kritsuk). We will also show that there are important
evolutionary processes within the disc itself on this longer
timescale.  In particular, while the initial stripping of the outer
disc occurs rapidly, we find that the internal disc instability
facilitates continued ablative sweeping.  Radiative cooling, or at
least the absence of compression heating is an important part of this
instability.  We present new analytic models to help understand this
multi-step stripping process.  We also examine the evolution of the
angular momentum of the gas disc, as another tool for understanding
the complex dynamics.

\section{STRIPPING THEORY AND A SIMPLE ANALYTIC MODEL}

Gunn and Gott (1972) assumed that gas in the bulk of the
disc was bound primarily to the local disc potential, and that in the
case of a face-on wind, removal of gas to a distance of one vertical
scale height was equivalent to stripping.  Abadi et al.\ updated this
criterion by adding bulge and halo gravity to the disc gravity, but
like Gunn and Gott they considered only the local balance between
gravity and ram pressure within the disc to derive a stripping radius.

In the case of a face-on wind impacting a disc bound by a massive
halo, the gravitational force of the halo does not oppose the wind
pressure initially, because the force vectors are perpendicular.  As a
gas element is swept vertically out of the disc (henceforth, the
z-direction), to a distance of several scale heights, the disc gravity
diminishes.  On the other hand, the projection of the radial halo
gravity vector into the z-direction, and opposite the ram pressure
increases.  This halo component can grow quite large, so that the
greatest resistance to sweeping can be found at some distance from the
disc.

In fact, because of this simple geometric effect, the halo gravity can
exceed the ram pressure force, and bind the gas to a region
substantially offset from the initial disc.  The region is like a
potential well that has been displaced by the addition of the ram
pressure force.  This effect, together with possible shadowing from
the ram pressure, allows gas elements to hang up at some distance
behind the original disc, rather than being completely removed.

The trajectories of some clouds in Kritsuk's (1983) models show that
they experienced this effect.  Indeed, the dashed curve in Figure 1 of
Kritsuk's paper would seem, by its definition, to be part of the
boundary of the displaced binding region.

\begin{figure*}
\includegraphics{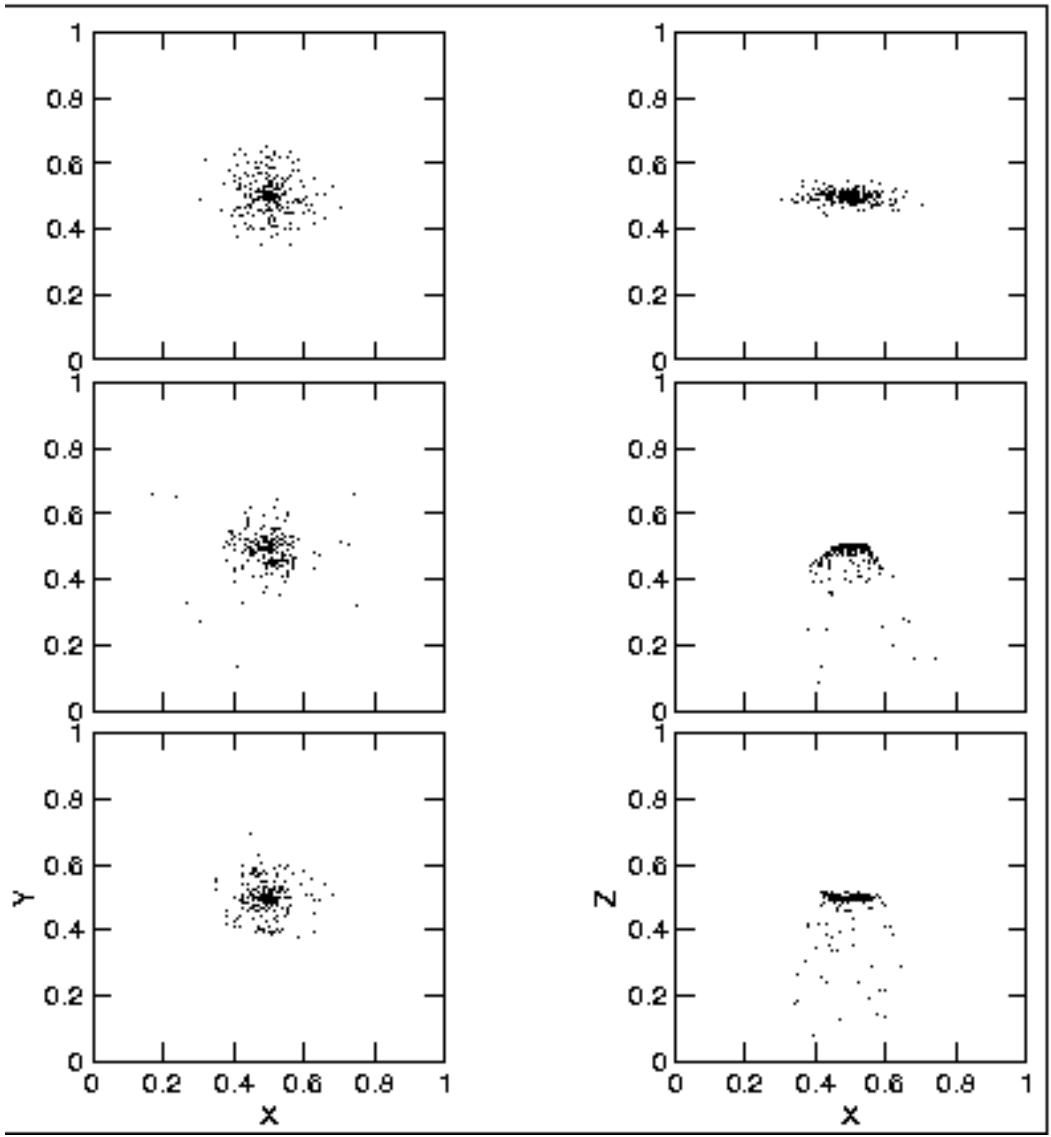}
\caption{Three snapshots with two views each of the fiducial model.
The top two plots are for timestep 0, where the relaxed galaxy is
placed in the ICM wind.  The middle two are for timestep 3159, and the
bottom two are for timestep 6000 ($t = 9.9 \times 10^7$ and $1.9
\times 10^8$ yrs., respectively, in the adopted scaling).  In this,
and similar figures that follow, only every fifth gas disc particle
has been plotted.}
\label{c1a}
\end{figure*}

\begin{figure*}
\includegraphics{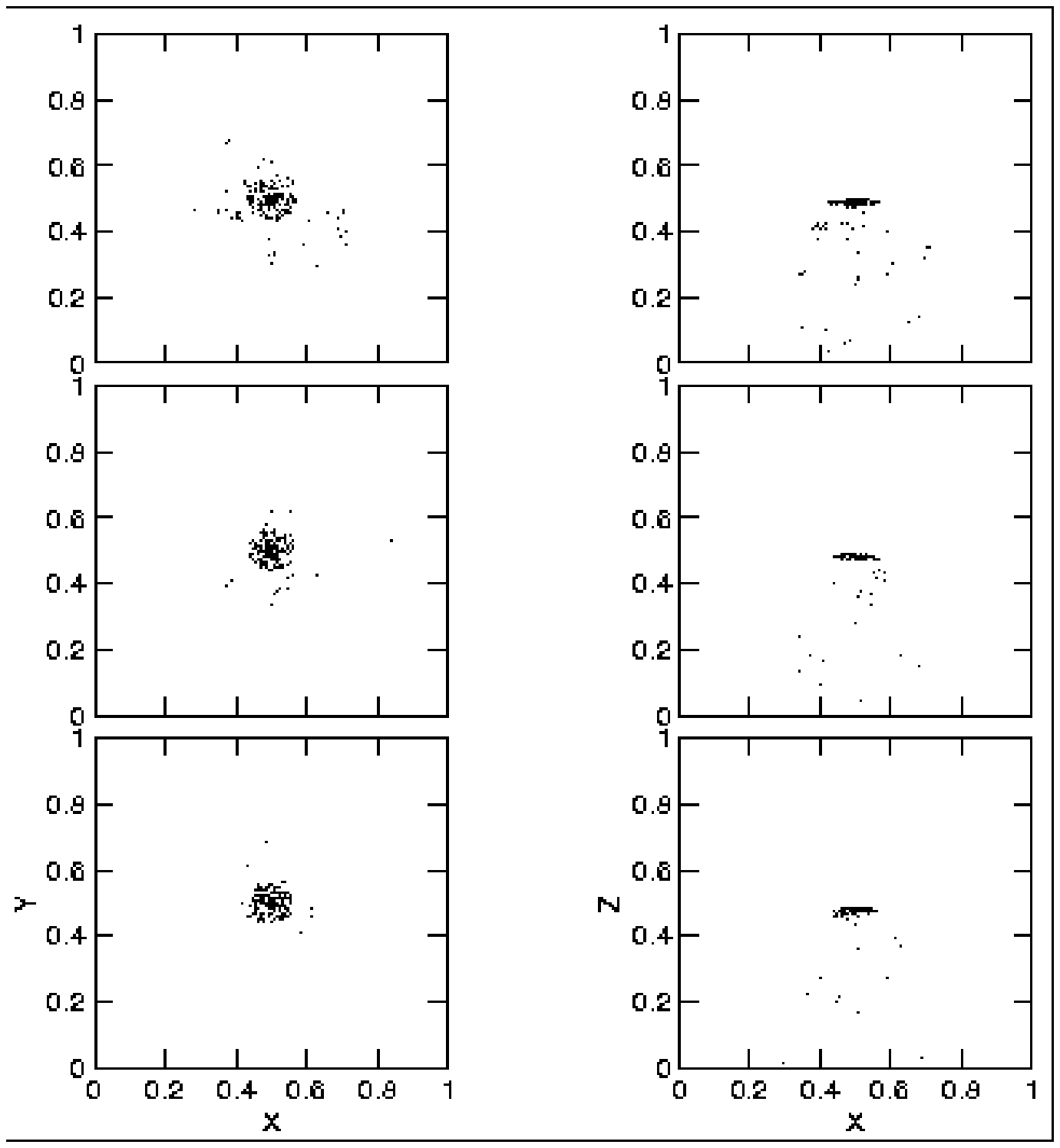}
\caption{Three late time snapshots with two views each of the fiducial
model.  From top to bottom the snapshots are at times of $3.8, 5.0,$
and $5.6 \times 10^8$ yrs., respectively.}
\label{c1b}
\end{figure*}

The following simple analytic model illustrates these effects.  Assume
the galaxy disc is face-on into the wind, with the wind flowing from
positive z values to negative.  Then the ram pressure force on a gas
element or cloud in the disc is,

\begin{equation}
F_{w} = \rho_{w}v_{w}^{2}A_{cl},
\end{equation}
 where $\rho_{w}$ is the mass density of the wind, $v_{w}$ is the
velocity of the wind in the z-direction, and $A_{cl}$ is the
cross-sectional area of a gas cloud in the galaxy.

For simplicitiy, assume that the halo is the only important source of
gravity.  Then the component of gravitational force on the cloud in
the z-direction is,

\begin{equation}
F_{h} = 
\frac{GM(r)m_{cl}}{r^2}\frac{z}{r} = \frac{m_{cl}v_{c}^2z}{r^2},
\end{equation}
where M(r) is the mass contained in a region of radius r, $m_{cl}$ is
the mass of the cloud, z is the vertical distance, and $v_c$ is the
local circular velocity.

In the second equality of equation (2) the halo mass interior to the
cloud radius has been formally eliminated in favor of the local
circular velocity, which we assume is constant in the region of
interest. We equate the two forces to derive not a ``removal radius'',
but rather a curve of r versus z, which when rotated around the axis
of symmetry gives the boundary surface of the displaced binding
region, as follows:

\begin{equation}
\frac{z}{r} = \frac{\rho_{w}rv_{w}^{2}A_{cl}}{m_{cl}v_{c}^2}.
\end{equation}
Generally, this an ovaloid region.

In terms of the initial orbital radius of the cloud in the disc, R,
this becomes 

\begin{equation}
\frac{zR}{r^2} =
\frac{\rho_{w}Rv_{w}^{2}A_{cl}}{m_{cl}v_{c}^2} = W(R),
\end{equation}
where $r^2 = z^2+R^2$.  Let $r' = r/R = 1/sin\theta$, $z' = z/R =
tan\theta$.  Then,

\begin{equation}
\frac{z'}{1+z'^2} = W.
\end{equation}

Solving for $z'$, we get,

\begin{equation}
z' = \frac{1}{2W}\pm[\frac{1}{4W^2} - 1]^{1/2}.
\end{equation}
we need $W < 1/2$ for a real solution, if $W > 1/2$, for a given R,
there is no binding region.  If $W < 1/2$ for some R, the cloud will
move to the bound region, and probably pursue a complex trajectory
determined by local pressure and shadowing forces.

Note that the $R/v_c$ is approximately equal to the local free-fall
time.  The remaining terms in the middle of eq. (4) can be written as
$(\rho_{w}v_{w}^{2}A_{cl})/(m_{cl}v_{c})$, which is the ram pressure
force on the cloud divided by the cloud momentum, i.e., the inverse
momentum change timescale. Thus, W is the ratio of the gravitational
restoring time, $\tau_{ff}$, to the momentum change time due to wind
pressure, $\tau_{mom}$.

\section{SIMULATION CODE AND INITIAL CONDITIONS}

\subsection{Simulation Code}
Our simulations were produced with the serial code Hydra 3.0, which
has been made publicly available by H.\ Couchman, P.\ Thomas, and F.\
Pearce.  The Hydra program implements smoothed particle hydrodynamics
(SPH) and calculates gravity with an adaptive particle-particle (PP),
particle-mesh ($AP^{3}M$) algorithm (for details see Couchman et al.\
1995, Pearce \& Couchman 1997).  For a typical timestep in our models,
adaptive refinements were carried out on about 1000 gas particles in
the grid center, with a scaling factor of 2.67.

The simulations were all run using an adiabatic equation of state.
Optically thin radiative cooling was calculated via the tables of
Sutherland and Dopita (1993), which were supplied with the Hydra code.
Cooling times were not used to limit the size of the computational
timestep, since the dynamical time is usually longer than the cooling
time.  Particles evolve adiabatically and are only cooled at the end
of a given timestep, at constant density.  

We note for completeness that the Sutherland and Dopita cooling curves
include atomic and ionic line and continuum processes for $T \ge 10^4
K$. These curves were also calculated at several metallicities, and
these alternate curves are included as options in the Hydra code.
However, all of our calculations were carried out at solar
metallicity.  In the range $T = 10^4 - 10^6$ the cooling is dominated
by H and He line and recombination cooling, and so, is relatively
insensitive to the metallicity.

Nonetheless, this is a very approximate treatment of the complex
thermal physics involved in the interaction of hot ICM gas and the
multi-phase disc gas.  Processes such as X-ray photoheating, molecular
cooling and the thermal effects of grain destruction are neglected.
However, these effects are most important at $T < 10^4 K$. At higher
temperatures it is likely that radiative cooling (e.g., near the peak
of the atomic cooling curve) dominates compressive and shock heating
in cool-to-warm gas elements originating in the galaxy disc, even when
these elements are swept out of the disc (see Section 5.1).  Our
treatment does capture this basic effect, and confirms simple rate
estimates. Several of the observational works cited in the
introduction have discovered extradisc HI gas, including long
filaments, apparently stripped from the cluster galaxy.  Abadi et al.
(1999) and Quilis et al. (2000) did not include cooling in their
models, arguing that it is unnecessary since their codes cannot
resolve molecular clouds.  Vollmer's (1999, 2000) sticky particle code
is intrinsically isothermal.  We agree that dense clouds cannot at
present be resolved, but our disc gas particles can represent diffuse
atomic gas and large cloud envelopes.  Several of our results on the
evolution and kinematics of stripped gas depend on cooling, as we
discuss below.

\subsection{Initial Conditions}

In each of the computational runs, we start with a model disc that
contains 29100 particles, divided up as follows: 9550 gas particles,
9550 disc star particles, and 10000 dark matter particles.  The masses
for each of the particle types, in code units, are: $3.0 \times
10^{-5}$ for gas and star particles, and $6.0 \times 10^{-4}$ for dark
matter particles.  The mass unit is $1.0 \times 10^{10} M_{\odot}$, so
each star or gas particle represents $3.0 \times 10^5 M_{\odot}$.
This gives a total galaxy mass of $6.57 \times 10^{10} M_{\odot}$, a
relatively small galaxy. 

The model galaxy was built first by initializing the halo particles in
an isothermal distribution, with an outer radius of about 0.3
grid units, and then run for several thousand steps to allow
relaxation.  Stellar and gas disc particles were initialized on
circular orbits with a small, three-dimensional random velocity
component added.  The model was relaxed with no ICM wind after the
addition of each component.

The model galaxy is placed in a grid ranging from 0.0 to 1.0 in units
for each direction, and centred at (0.5,0.5,0.5) in (x,y,z).  The code
unit of length is about equal to 100 kpc.  The gas disc has a radius
of 15 kpc.  The radius of the stellar disc is about 10 kpc. in x and
y, and the dark halo is spherical with a radius of about 10 kpc.

When this initial galaxy is read into Hydra, it is surrounded with
80000 gas (ICM) particles.  These particles were randomly dispersed in
50 layers of 1600 particles each.  The layers were spaced 0.0196 units
apart (1.96 kpc.) in the z-direction, so as to create a uniform ICM.
This ICM particle spacing is also comparable to the cell size of the
$64^3$ large-scale grid.  As will be seen below, these initial
conditions do allow us to (minimally) resolve the large-scale bow
shock around the galaxy. Each of the ICM particles was given a mass
10\% of that of a gas particle belonging to the galactic disc.  In
using this reduced mass we follow the recommendations of Abadi et al.\
(1999), who point out that if the mass is too high, the ICM particles
will punch through the galaxy disc.

In these models the temperature of the initial ICM was set to 5.0
units, of $4.6 \times 10^5$ K, so $T = 2.3 \times 10^6$ K.  In
addition, the density of the wind is approximately $7.3 \times 10^{-5}
{cm}^{-3}$.  These values are typical for a moderate galaxy cluster.

\begin{figure}
\includegraphics{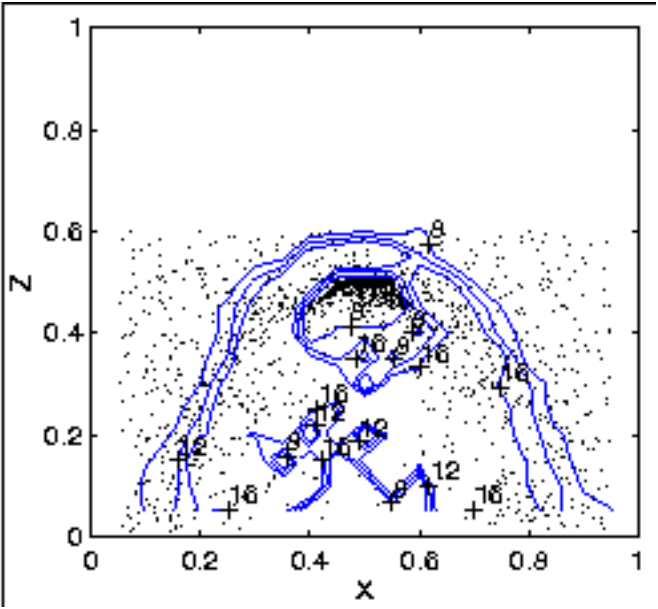}
\caption{Wind particles and temperature contours of 8, 12, and 16
units (corresponding to temperatures of $3.7, 5.5, \& 7.4 \times 10^6$
K) for the fiducial model at $t = 9.9 \times 10^7$ yrs.  The ICM bow
shock and reflection shock around the gas disc are evident in these
contours.  Contours in the rarefied wake are very uncertain.  Only
particles with $0.425 \le y \le 0.575$ grid units are shown.}
\label{tshock1}
\end{figure}

In order to simulate the galaxy traveling through the ICM, we kept the
galaxy stationary, while having the intracluster plasma flow past the
galaxy as an effective wind.  This flow was directed from +z to -z.  A
control simulation was also run with no wind, for a time equal to that
of the other runs.  No discernable change was seen in the model galaxy
in this run, including no spiral or bar instabilities.

\subsection{Boundary Conditions}

The following boundary conditions were employed. First, in the x, y,
and z directions, the particles are required to remain between 0.01
and 0.99 in code units.  These conditions were checked every few
timesteps.  This is done to maintain efficient computing performance
while still ensuring that few particles were lost from the grid (in a
typical run, less than 0.3\% of the intracluster gas particles).  In
addition, when particles reach the lower bound of $z= 0.01$, they are
randomly placed back into the top of the ICM flow, and given a mass
1/10 that of the gas particles in the galactic disc.

Thus, some of the particles put back into the ICM were formerly gas
particles from the disc.  When they are brought back into the grid,
they are given the smaller ICM particle mass.  So, in determinations
of which gas disc particles are still on the grid, these particles are
not included (for further details see Schulz (2000)).

\section{MODEL RESULTS}

In this section, we present the results of our new simulations, with
further analysis in the following sections.  We begin with a face-on
'fiducial' model.  Next we consider models in which the galaxy is
tilted relative to the ICM wind, then a model with a larger wind flow
velocity, and finally models in which the temperature of the disc gas
was initially set to a high temperature, but allowed to cool, to
crudely model the effects of a strong global starburst.  The
parameters of each of the data runs are shown in Table \ref{ov}.

\begin{table*}
\caption{Overview of the data runs.}
\label{ov}
\begin{tabular}{ccccc}\hline
Run \# & Vflow (km/s) & $\rho$ ($10^{-5}$ ${cm}^{-3}$) & Secondary Heating 
($10^6$ K) & Tilt\\[1.0ex] \hline \\
1 & -1000 & 7.3 & -- & $0^{\circ}$\\
2 & -2000 & 7.3 & -- & $0^{\circ}$\\
3 & -1000 & 7.3 & 1.4 & $0^{\circ}$\\
4 & -1000 & 7.3 & 3.5 & $0^{\circ}$\\
5 & -1000 & 7.3 & -- & $40^{\circ}$\\
6 & -1000 & 7.3 & -- & $60^{\circ}$\\
7 & -1000 & 14.6 & -- & $0^{\circ}$\\
\\
\hline\\
\end{tabular}
\end{table*}

\subsection{Fiducial Model}

The fiducial model is a face-on case with a wind velocity of -977 km/s
with the adopted scaling.  We ran the code for 18000 timesteps,
equivalent to a time span of $5.64 \times 10^8$ yrs. with this
scaling.  At -977 km/s, a particle starting at the top crosses the
entire grid over 5 times by the end of the run.

\begin{figure*}
\includegraphics{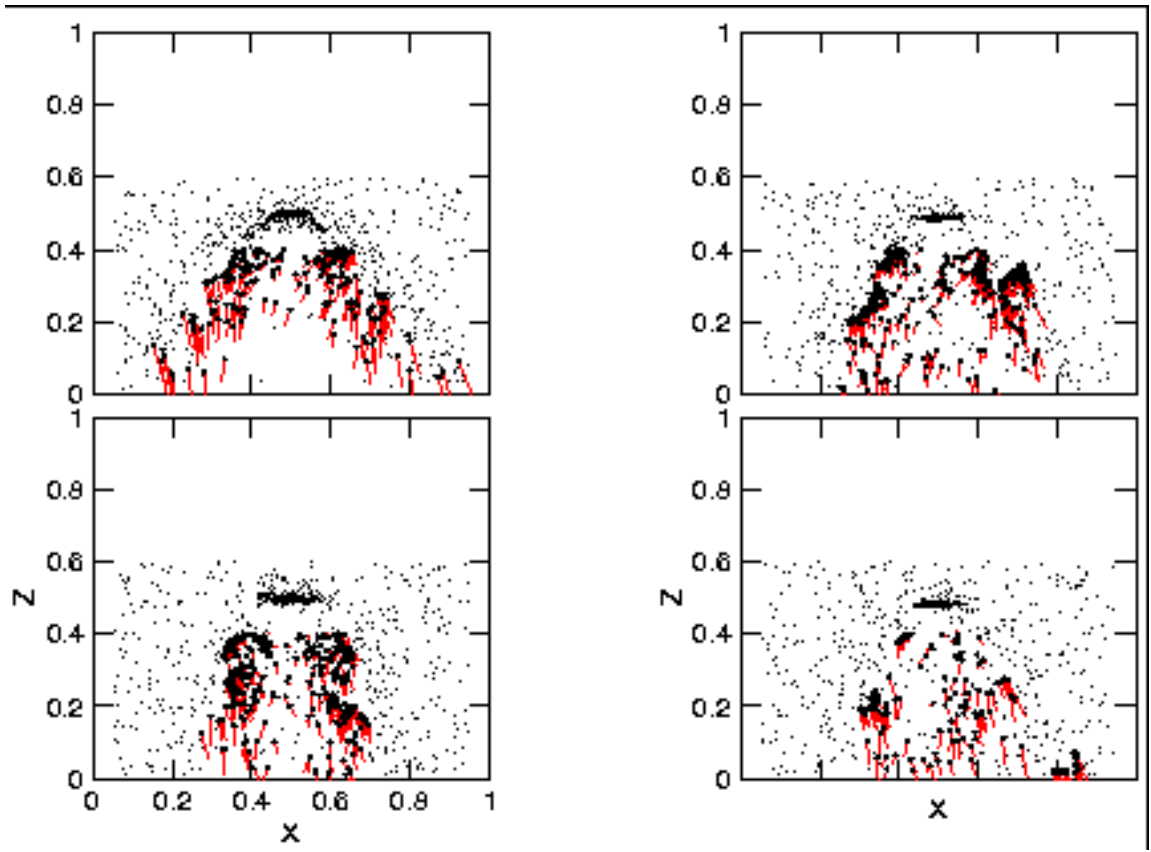}
\caption{X-Z Velocity vector components for those particles stripped
from the gas disc which are at or below 0.4 grid units in the
z-direction.  The dots are the bases of the vectors.  Timesteps shown
are $0.99, 1.9, 3.8, and 5.0 \times 10^8$, from the upper left to
lower right panels.  The low velocities, and even occasional upward
velocities, at and just below the plane z = 0.4, illustrate the
hang-up and fall-back phenomena.}
\label{vp1a}
\end{figure*}

\begin{figure*}
\includegraphics{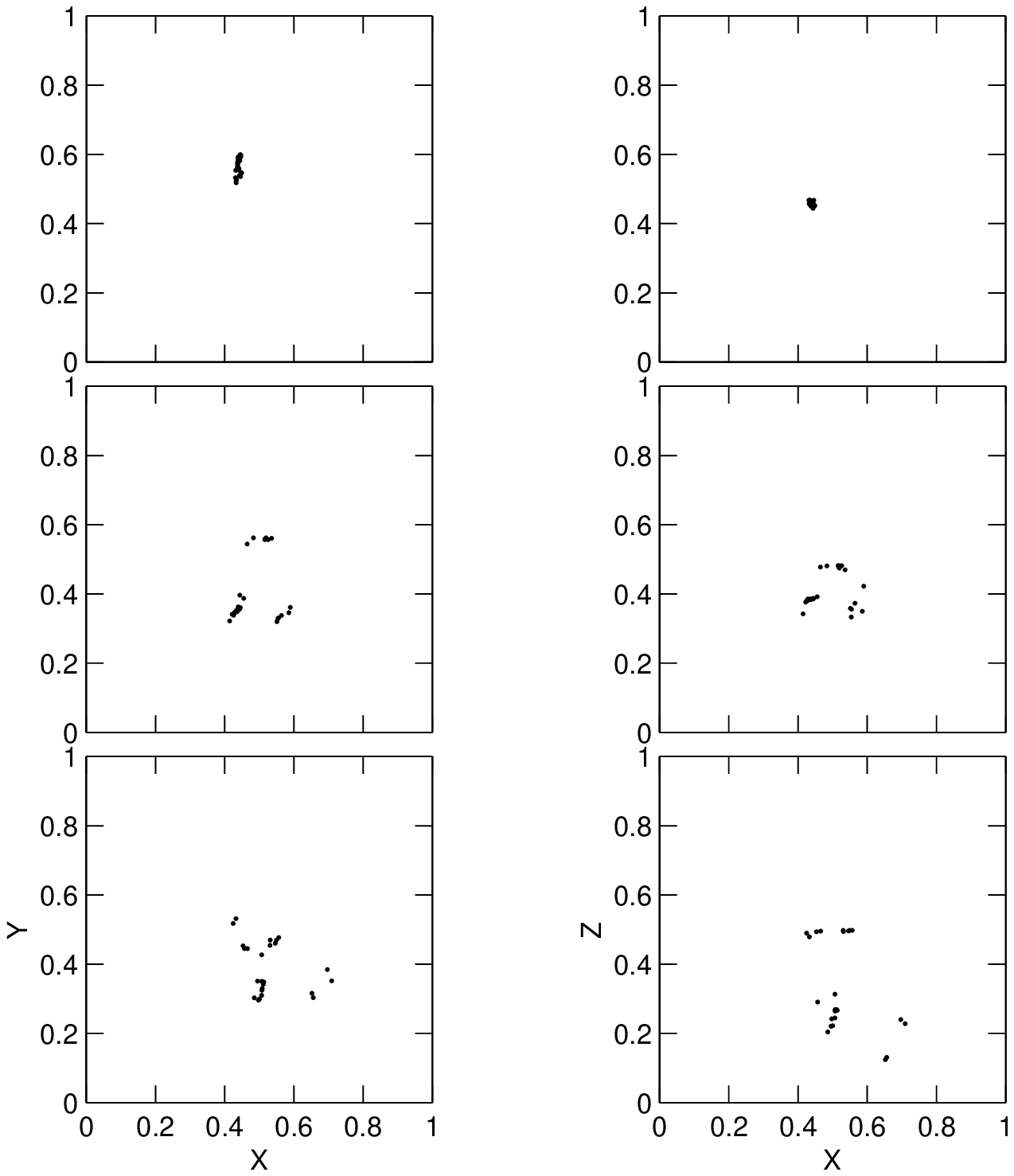}
\caption{Two orthogonal views of the evolution of a representative
(unbound) cloud of gas particles at three times in the fiducial model.
The timesteps shown are the same as in Figure \ref{c1a}. See text for
details.}
\label{fila}
\end{figure*}

In Figures \ref{c1a} and \ref{c1b}, we show the particles from the gas
disc (only) at various timesteps over the evolution of the model.
These figures bear a qualitative resemblance to Figure 12 of Farouki
\& Shapiro (1980).

During the early evolution of the model (Figure \ref{c1a}), the outer
disc deforms rapidly.  Mass loss begins as particles move towards the
bottom of the grid.  By the time of the second snapshot spiral waves
appear in the outer disc, and a ring of compressed gas separates the
inner and outer discs.  This ring also divides the planar and
nonplanar regions.

A clear bow shock also forms early in the ICM.  This is shown in
Figure \ref{tshock1} after about $10^8$ yrs.  For clarity, wind
particles above 0.6 code units in z have been removed, and only a
section ranging from 0.425 to 0.575 units in y is shown in Figure
\ref{tshock1}.  The opening angle of the shock is quite broad (see
discussion below), so that the shock extends across most of the x-axis
at the bottom of the grid.  Very few particles are visible behind the
bow shock, signifying that few, if any, wind particles are able to
pass through the galaxy unimpeded.  The bow shock visibly compresses
the disc in the vertical direction.

The temperature contours for the bow shock are also shown in Figure
\ref{tshock1}.  The numbers are in the code units described in Section
3.2.  Postshock heating, as well as the heating around the galaxy disc
gas by the reflected shock, is visible. Radiative cooling keeps gas
within the disc from heating to these million degree temperatures.

Figure \ref{c1b} shows that mass loss tapers off at the end of the
run.  Low-density material is removed first.  This consists of gas
found between spirals or in low-density spirals.  Higher density
spirals are stripped later, and then tend to ``hang up,'' or in some
cases, even fall back onto the disc.  These phenomena are predicted by
the analytic model, and can be seen in Figure \ref{vp1a}.  The
filamentary appearance of the stripped spirals is reminescent of HI
filaments observed in several galaxies, as described above.

In particular, the last x-z plot in Figure \ref{c1b} also shows the
existence of a tail of gas extending out beyond the galaxy a distance
of at least 20kpc.  The development of this tail from a disc can be
discerned in the preceding snapshots.

The shock angle, as measured relative to the bottom of the grid,
varies over time (ranging from 60 to $75^{\circ}$).  These changes are
due not only to stripping but to vertical and radial oscillations of
the disc following the appearance of the bow shock.  These changes can
most easily be seen in the position of the shock at the bottom of the
grid in images like Figures \ref{c1a}, \ref{c1b}, and \ref{vp1a}.

In order to better understand the evolution of stripped particles, we
examined small, arbitrarily defined gas clouds (groups of particles)
at various times.  Each of the clouds that we looked at was identified
at 3159 timesteps and followed thereafter.  The motions of individual
clouds are quite complex and dynamic.  An example of such a cloud is
shown in Figure \ref{fila}.  Initially, it consisted of 129 gas
particles.  By 18000 timesteps, only 42 were left in the grid.  Of the
particles that were left, most appear to have been accreted back onto
the disc.  This is, however, a rather exceptional example.  Most
clouds in the outer disc are eventually entirely removed.

\begin{figure}
\includegraphics{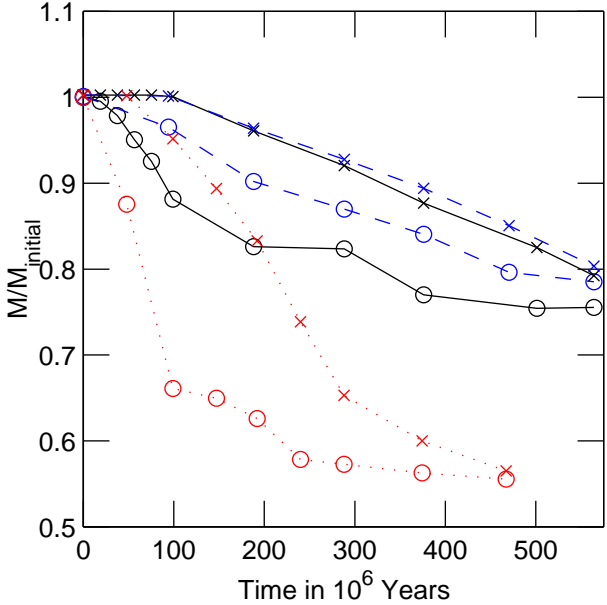}
\caption{Comparison of mass loss in three models.  The circles
connected by the solid line segments represent the gas in the fiducial
model retained in a rectangular volume around the original gas disc as
a function of time (see text for details of the box bounds). The
circles connected by dashed and dotted lines represent the gas
retained within the same box in the $40^{\circ}$ tilt and $2000 km/s$
models, respectively. The solid, dashed and dotted curves connecting X
marks depict gas mass remaining on any part of the computational grid,
in the fiducial, $40^{\circ}$ tilt and $2000 km/s$ models,
respectively.}
\label{m1}
\end{figure}

We examined both the mass entirely removed from the computational
grid, and the mass lost from a fixed volume enclosing the original
disc, as a function of time (see Figure \ref{m1}). This volume is
defined as a rectangle that goes from 0.3 to 0.7 grid units in x and
y, and 0.425 and 0.575 units in the z-direction.  The figure shows
that the disc lost the most material within the first $10^8$ yrs.
This result is roughly consistent with that of Abadi et al.\ (1999).

After about 375 million years only a few particles are still being
stripped off the disc.  On the other hand, after 100 million years,
the rate of mass loss from the grid as a whole remains fairly
constant.  The difference between the two curves lies in the fact that
some particles linger in the grid for long timescales, while other
particles are slowly accelerating away from the effective potential
minimum.

\begin{figure}
\includegraphics{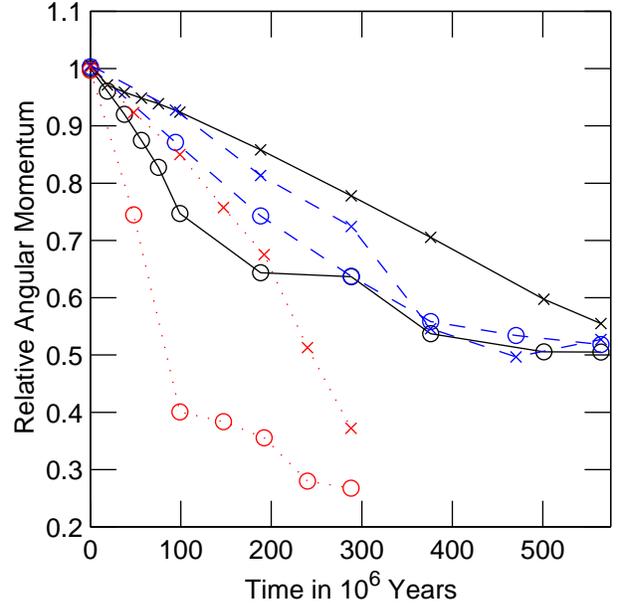}
\caption{Comparison of angular momentum loss in three models.  The
circles connected by the solid line segments represent the gas angular
momentum in the fiducial model retained in a rectangular volume around
the original gas disc as a function of time (same box as in previous
figure). The circles connected by dashed and dotted lines represent
the gas angular momentum retained within the same box in the
$40^{\circ}$ tilt and $2000 km/s$ models, respectively. The solid,
dashed and dotted curves connecting X marks depict angular momentum
remaining on any part of the computational grid, in the fiducial,
$40^{\circ}$ tilt and $2000 km/s$ models, respectively.}
\label{am1}
\end{figure}

Figure \ref{am1} is a plot of total angular momentum versus time for
particles that originated in the gas disc.  The falloff in angular
momentum over time appears quite similar to that of the gas mass.  The
ratio of mass loss to angular momentum loss rises rapidly within the
first 100 million years but then settles down to a value of
approximately 0.5, meaning that about twice as much angular momentum
is being lost as gas mass.  This is because the material with high
specific angular momentum spreads the most, and thus, is easiest to
strip.  This high rate of angular momentum loss accounts for most of
the compression we see in the disc with time.

We found, somewhat to our surprise, that the stellar disc, too,
appears to be affected by the intracluster medium.  While the stellar
disc was not altered in size or shape, by the end of the run it had
been displaced roughly 0.02 length units in the negative z-direction,
which translates to $6.17 \times 10^{16}$ km, or 2 kpc.  We believe
that this displacement is an indirect result of the bulk displacement
of the gas disc which is not stripped.  The mutual gravity of the gas
and star discs, which are of equal mass in our models, prevents them
from separating, and thus, pulls the stellar disc down relative to the
fixed grid.

We estimate that the net kinetic energy that is transferred from the
wind to the gas disc exceeds the disc self-gravity.  Thus, it is
adequate to account for the displacement.

The effect provides an interesting, if rough, check on the numerical
model.  Since the outer disc will be most affected by the wind force,
we consider the vertical wind force acting an annular region in the
outer part of the disc.  That is, we essentially view the outer disc
as analogous to a solar sail which is dragging its payload, where the
payload is the stellar disc.  Vertical force balance is achieved when

\begin{equation}
\rho_wv_{w}^2A_{ann} = \frac{GM(r)m_{ann}}{r^2}\frac{z}{r}.
\end{equation}
Here, $\rho_w$ is the wind mass density, $v_w$ is the wind velocity in
the z-direction, $A_{ann}$ is the area of the annular region in x and
y, M(r) is the disc mass contained in r, $m_{ann}$ is the mass of
particles in the annulus, r is the distance from a particle in the
annulus to the potential center, and z is the vertical displacement of
the disc.

We can substitute in $v_c^2/r$ for $GM(r)/r^2$, where $v_c$ is the
effective centripetal velocity.  We find that,

\begin{equation}
z = \frac{\rho_w v_w^2 A_{ann} r^2}{m_{ann}v_c^2} = 
\frac{\rho_w v_w^2}{\rho_{ann} v_c^2}  
\biggl(\frac{r}{2h}\biggr)^2 2h, 
\end{equation}
where to obtain the last form we use the relation $\rho_{ann} =
A_{ann}(2h)$, with disc thickness, 2h.  With values from the fiducial
model, we calculate that $z \simeq h$ in code units.  Since $h \simeq
0.02$, this estimate agrees well with the numerical result.  The
displacement of the disc is greater in a higher flow velocity model,
as the above equation predicts.

\begin{figure*}
\includegraphics{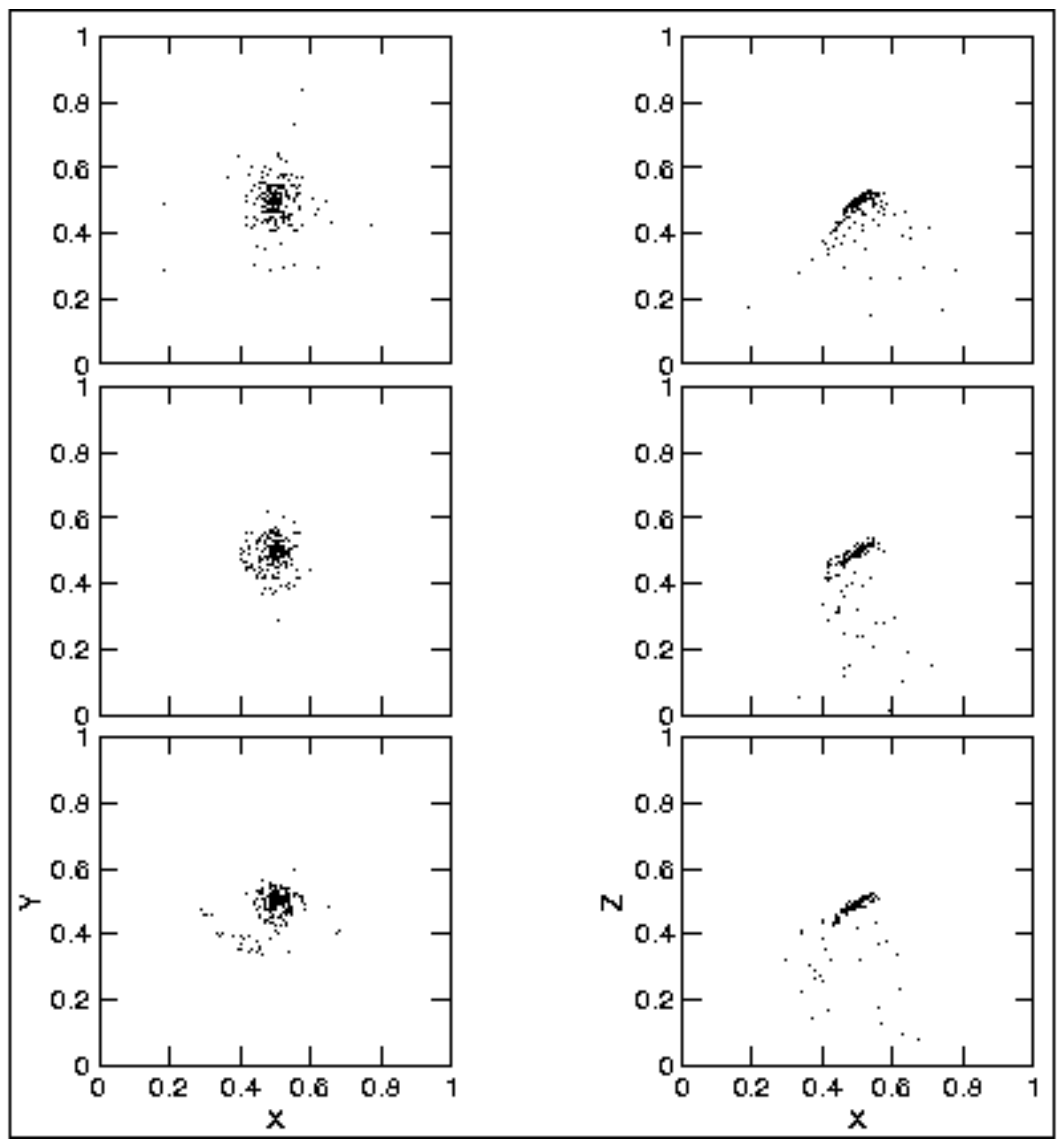}
\caption{Galaxy tilted $40^{\circ}$ about the z-axis (the wind
direction).  From top to bottom the snapshots are at times of $0.94,
1.9,$ and $2.9 \times 10^8$ yrs., respectively.}
\label{c4a}
\end{figure*}

\begin{figure*}
\includegraphics{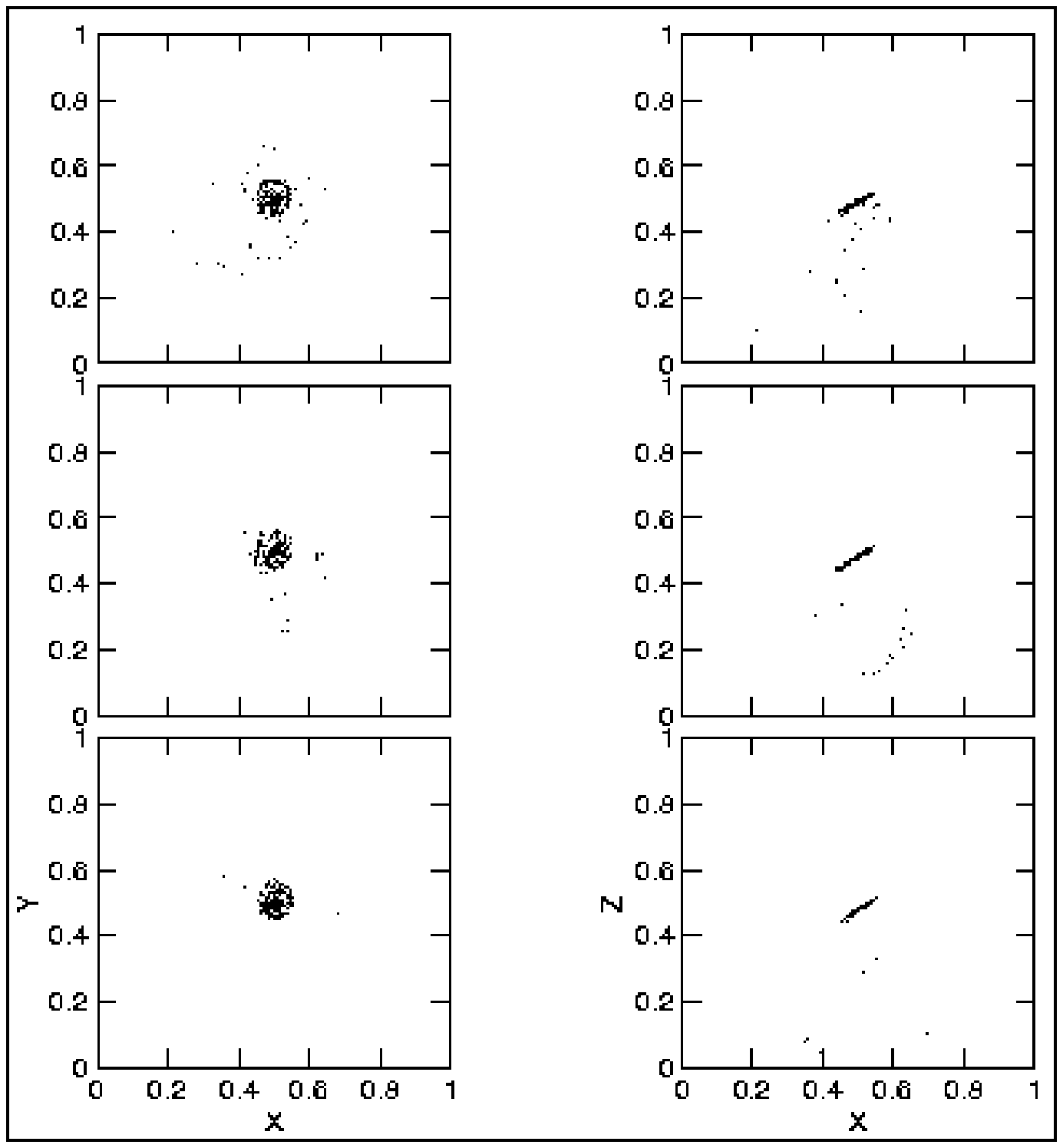}
\caption{Galaxy tilted $40^{\circ}$ about the z-axis, at late
times. From top to bottom the snapshots are at times of $3.7,
4.7,$ and $5.6 \times 10^8$ yrs., respectively.}
\label{c4b}
\end{figure*}

\subsection{Tilted Models}

Next we describe two simulations run to study the effect of
inclination on gas sweeping.  In the first one, the galaxy was tilted
by $40^{\circ}$ from the z-axis.  Six timesteps for this run are
displayed in Figures \ref{c4a} and \ref{c4b}.  One noteworthy aspect
of these plots is that, after 18000 timesteps ($5.6 \times 10^8$ yrs.),
almost all of the gas from the disc is either still in the disc or has
been removed from the grid.  Much less remains between the gas disc
and the bottom of the grid than in the fiducial model.  Another
interesting feature is the appearance of a large ``tail'' of material
at timestep 9200 ($2.9 \times 10^8$ yr.).  It includes particles
ranging from 0.35 to 0.7 units in x, 0.25 to 0.5 in y, and 0.2 to
0.42 units in z.  From step 9200 on, it can be seen moving down and to
the right in the x-z plots.  At step 18000, only a small portion of
the original tail remains in the region from 0.6 to 0.8 in x, and from
0.0 to 0.15 in z.

This tail can first be seen at step 6000 as a strong spiral arm in the
x-y view of Figure \ref{c4a}.  Subsequent views in the two projections
reveal that as the arm is rotated into a position where it is moving
against the wind, it is stripped. This seems to be an example of the
kind of rotationally aided stripping postulated by Phookun \& Mundy
(1995), and mentioned in the introduction.
  
\begin{figure}
\includegraphics{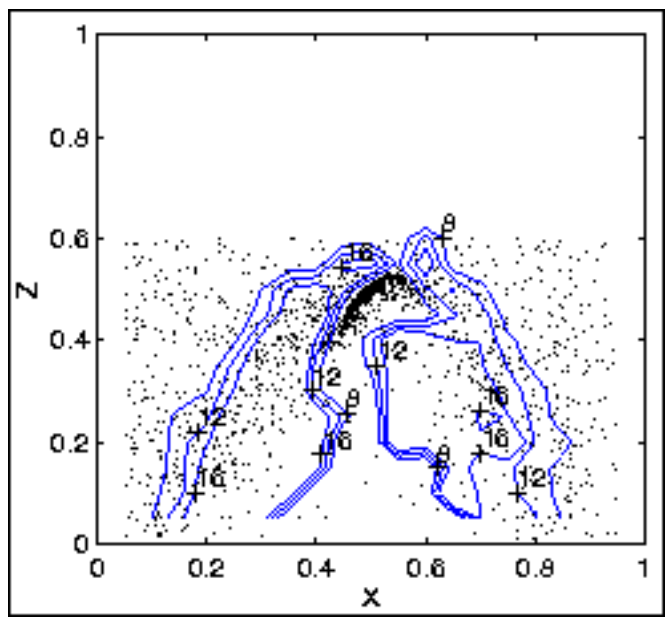}
\caption{Galaxy tilted $40^{\circ}$ about the z-axis.  Temperature
contours of 8, 12, and 16 units (corresponding to temperatures of
$3.7, 5.5, \& 7.4 \times 10^6$ K) at $t = 0.94 \times 10^8$ yrs.  Only
particles with $0.425 \le y \le 0.575$ grid units are shown.  Compare
to Figure \ref{tshock1}}
\label{tshock3}
\end{figure}

Figure \ref{tshock3} shows the temperature contours for the bow shock.
Although the temperatures are comparable to that of the fiducial
model, the contours show that the heating is much less symmetric than
before. Note, however, that the contouring was performed with a
coarser mesh resolution than the simulations themselves. Granted this,
and the limited particle numbers immediately downstream from the disc,
the contours there are uncertain.

\begin{figure}
\includegraphics{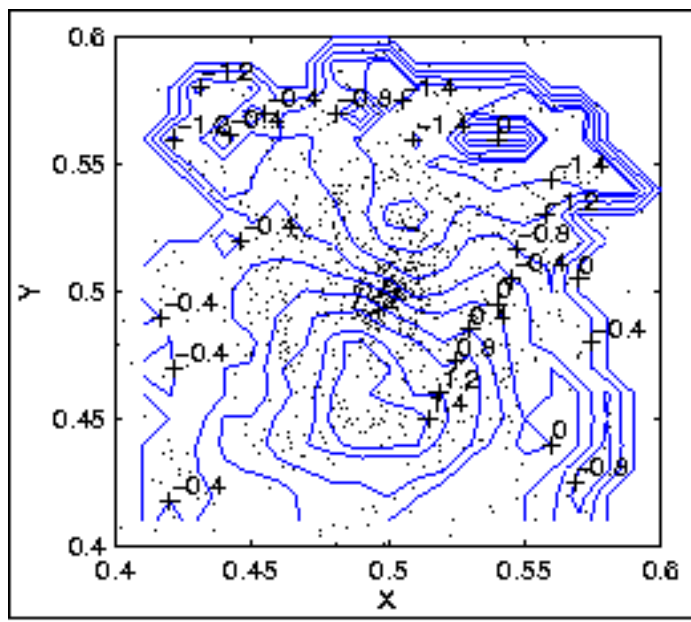}
\caption{Galaxy tilted $40^{\circ}$, at $t = 0.94 \times 10^8$ yrs.
Contours of z component of the velocity of the gas disc particles,
ranging from -1.6 to +1.2 (or -156. to 117. km/s).}
\label{vzcont}
\end{figure}

\begin{figure}
\includegraphics{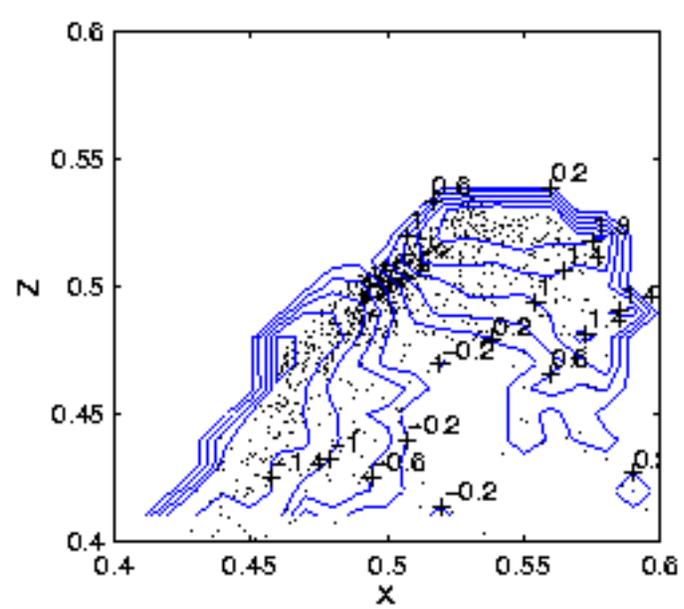}
\caption{Same as Figure \ref{vzcont}, except contours
of the y velocity ranging from -176. to +176 km/s.}
\label{vycont}
\end{figure}

Figure \ref{vzcont} shows the z-velocity contours on the x-y
projection of the disc at $10^8$ yrs. into this run.  In the inner
disc, within the ring, we see a more or less normal pattern of
projected rotation, with only small asymmetries.  Outside the ring,
the asymmetries are large.  At the top of the figure the disc rotates
with the wind.  There the velocities reach high amplitudes, and the
contours are compressed.  At the bottom the rotational flow opposes
the wind, and velocity contours are stretched out.  There is a general
similarity to the kinematic observations of NGC 4848 presented in
Vollmer (2000).  The kinematic continuity between disc and stripped
gas, which has been noted by a number of observers is also apparent in
Figure \ref{vzcont}.  This is also apparent in Figure \ref{vycont},
which shows contours of y velocities in the x-z plane.

Overall, roughly the same number of particles were removed in this
model as in the fiducial model (about 1000 particles over about 7600
timesteps).  Figure \ref{m1} shows that the initial mass loss is not
as rapid as the fiducial model, but by the end of the run, it is
comparable.

The total angular momentum loss over time is shown in Figure
\ref{am1}).  It is interesting that after about $3.75 \times 10^6$
yrs., the quantity around the disc is actually larger than that on the
whole grid!  The explanation for this is that a large cloud of
particles (the tail noted above), was counter-rotated, and then
stripped, as a result of an interaction with the ICM.
 
In the second tilted simulation, the galaxy was inclined $60^{\circ}$
about the z-axis.  Generally this model is similar to the previous
one.  However, the increased tilt narrows the bow shock.  By the end
of the run, a large part of the gas disc has settled into the
ring-like structure that we have seen in the previous models.  A
one-arm spiral develops, and is stripped much as in the previous run.
About 500 fewer particles have been removed than when the galaxy was
inclined $40^{\circ}$.

\subsection{Increased Flow Velocity Model}

Next we describe a simulation with twice the flow velocity of the
previous ones (-1953 km/s).  Like the others, this model was run for
18000 timesteps.  With the usual Courant-Friedrichs-Lewy (CFL)
condition (see e.g., LeVeque et al.\ 1998, 278) the numerical timestep
was reduced to about half its previous value.  Thus, the total run
time was about 288 million years.

\begin{figure*}
\includegraphics{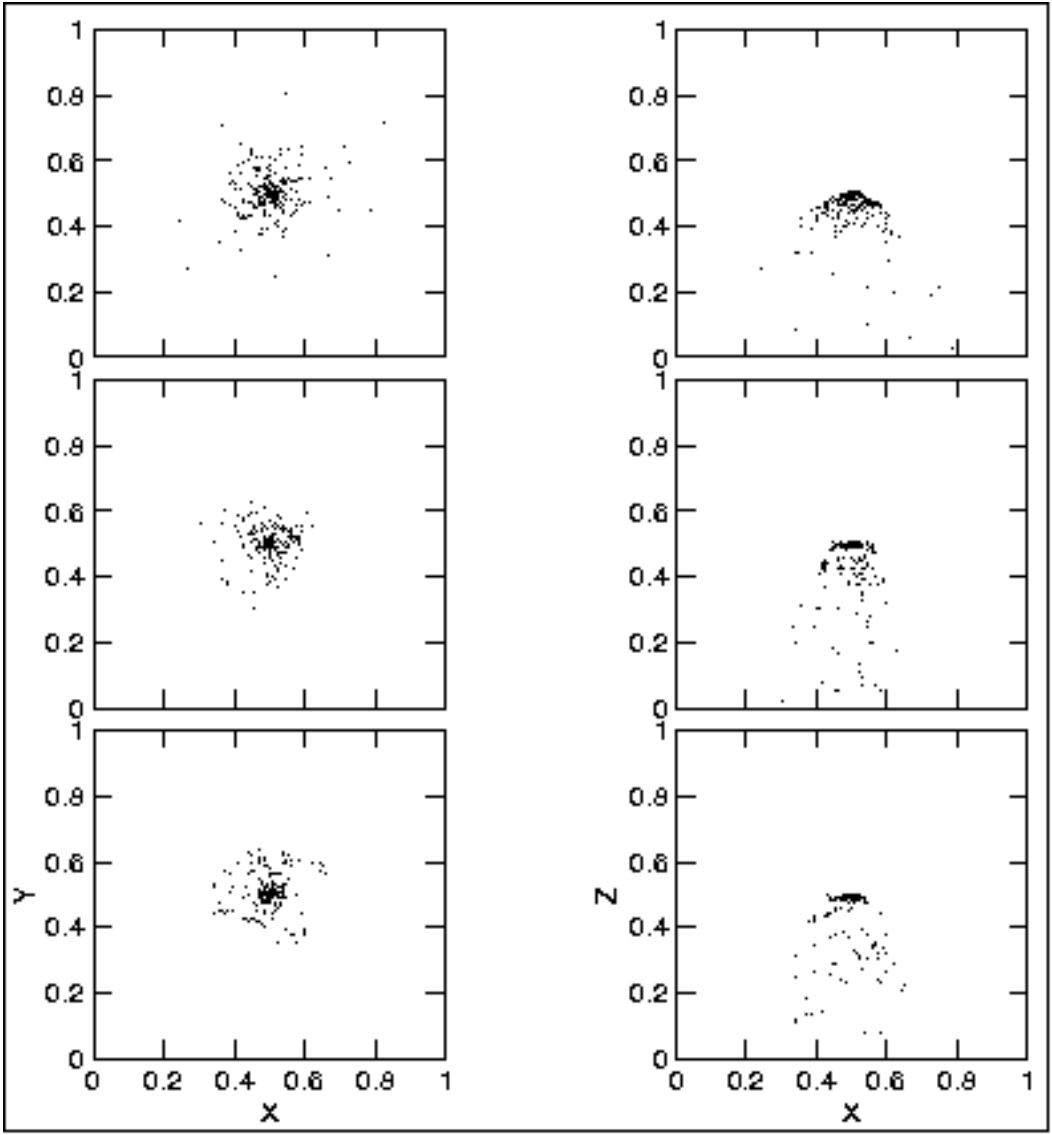}
\caption{Model with wind velocity $\approx$ -2000 km/s.  From top to
bottom the snapshots are at times of $0.47, 0.9,$ and $1.5 \times
10^8$ yrs., respectively.}
\label{c2a}
\end{figure*}

\begin{figure*}
\includegraphics{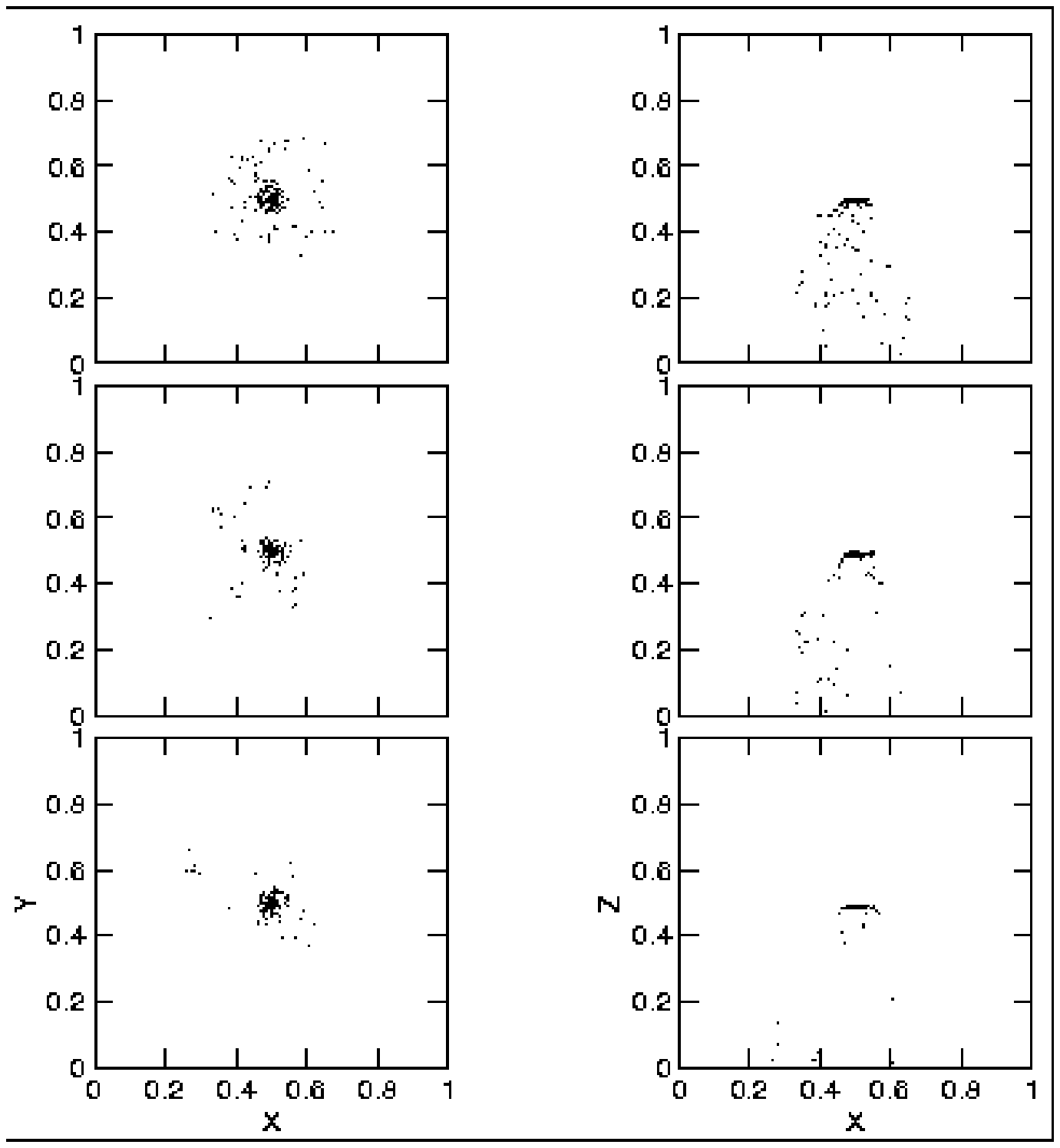}
\caption{Model with wind velocity $\approx$ -2000 km/s, at late
times. From top to bottom the snapshots are at times of $1.8,
2.4,$ and $2.8 \times 10^8$ yrs., respectively.}
\label{c2b}
\end{figure*}

Figures \ref{c2a} and \ref{c2b}) show snapshots of the gas particles
from the disc.  Although the plots are quite similar in appearance to
that of the fiducial model, the sweeping process occurs on a much
shorter timescale.

About 1000 more particles were lost in this case than in the fiducial
model, and in about half the time, see Figure \ref{m1}.  Here, the
mass loss around the disc is much more rapid within the first 100
million years, but beyond that it essentially levels off.  The ICM
parameters of this run are close to those of Run C of Abadi et
al. (1999, though our model galaxy is about 4 times less massive), and
the resulting mass loss curve is also very similar.  The total angular
momentum loss is plotted in Figure \ref{am1}.

Higher wind velocity produces more shock heating, and thus, higher
temperatures than before.  In this model they range from 14.7 million
K to 29.5 million K.

We also observed the motions of individual clumps of particles over
time in this model.  For nearly every one of the clouds we looked at,
all of the particles were removed from the grid by the end of the
simulation.  This is consistent with the sweeping criterion, in terms
of the W parameter, discussed above.  W is proportional to the square
of the wind velocity, so by doubling the velocity for this model, we
have effectively multiplied W values by 4, increasing the likelihood
of stripping.  More precisely, the volume of the ``trapping region''
in the halo, with $W < 1/2$, is very small.

\subsection{Models with Secondary Heating}

We also tested the effect of impulsive internal heating in the galaxy
disc, for example, heating as a result of a global burst of star
formation.  Specifically, we carried out two simulations in which the
gas disc of the fiducial model was impulsively heated after 6000
timesteps.  All other parameter values were kept the same as in the
fiducial model.

In the first case, the gas disc was heated to approximately 1.4
million K.  However, this temperature was still low enough that the
galaxy was able to cool rapidly, and there were no noticeable
differences in gas removal from the fiducial model.

In the second case, the gas disc was heated to about 3.5 million K.
This heat was not immediately radiated, and the gas particles in the
galaxy puffed out vertically over a timescale of less than 94 million
years.  By a time of $2.9 \times 10^8$ yrs., 1100 more particles
(11\%) were removed than in the case without heating.  After that, the
effects were the same overall as in the fiducial model. We conclude
that enhanced star formation may lead to only moderately enhanced
stripping on cluster crossing timescales, and substantial (or
prolonged) heating is required.

\section{DISCUSSION}
\subsection{The Deterministic, Multistage Sweeping Process}

There is a great deal of consistency in the results of the different
runs described above.  It appears that the stripping/sweeping process
proceeds through the same distinct stages regardless of galaxy tilt,
or ICM velocity and density, at least over a range of an order of
magnitude in these parameters. Based on the model results above and
previous work, we suggest that the process has the following
characteristic stages.

(1) {\it{Prompt stripping}} of the outer disc gas from the disc plane.
The effectiveness of this Gunn/Gott process has been emphasized in the
studies of Abadi et al. (1999) and Quilis et al. (2000).  Promptly
stripped material is lightly bound to the disc plane, but it remains
bound to the dark halo initially.

(2) Specifically, the stripped gas {\it{hangs up}} near the minimum of
the effective potential downstream from the disc, as described in
Section 2.  Thus, stripping from the disc is not always equivalent to
sweeping out of the galaxy.

Cooling plays an important role in the hang-up phenomenon.  Warm
clouds of a given mass, in a constant pressure environment, will have
a smaller cross section than an equal mass of hot gas, $A_{cl} \propto
\rho_{cl}^{-2/3} \propto T_{cl}^{2/3}$. In our models most of the
hung-up gas has a temperature of about $10^4 K$. While it is heated by
interaction with the ICM, this heating is not sufficient to push it
over the peak of the cooling curve. 

(3) Simultaneous with the first two stages is the {\it{formation of a
bow shock}} in the ICM, and also of a reflected shock around the gas
disc.  Until recently models have not provided much information about
the three dimensional structure of the shocks, although high
resolution two dimensional models, like those of Stevens et
al. (1999), are very suggestive.

The figures in Abadi et al. (1999) show the presence of bow shocks,
the model galaxies of Quilis et al. (2000) have a large central hole,
and as a result an annular bow shock forms with wind flow through the
centre.  These latter authors also contend that the wind flows through
kiloparsec scale holes in the disc ISM, and that this enhances prompt
viscous stripping.  We find a similar effect in our models, in the
efficient stripping of gas between dense spirals in the outer disc.
As described in Section 4, we also find that the bow shock expands and
contracts on large scales, following changes in the cross section of
the gas disc (which are due to oscillations, stripping, and angular
momentum evolution).

Despite holes and gaps, the global bow shock still forms in all the
simulations above, and compresses the disc.  Given the fact that the
clumpy, multiphase nature of the ISM is not fully resolved, we might
wonder if this result is correct.  E.g., if all the gas is promptly
swept except the cores of the largest clouds, then a global shock
would probably not form.  However, there are several reasons to think
that this will not occur.  Firstly, Crosthwaite, Turner \& Ho's (2000)
recent high-resolution (about 350 pc) HI map of the nearby spiral IC
342 shows that almost all of the disc is covered with diffuse,
resolved atomic gas with a column density of $ \ge 3 \times 10^{20}
cm^{-2}$, which is substantial.  Studies of the diffuse ISM in
galaxies suggest that this component is also enveloped in hotter
ionized or partially ionized phases of significant column density
(e.g., the review of Reynolds 1996 on this component in the Milky
Way). In addition the whole is threaded and connected by magnetic
fields.  Thus, it appears that the column density of the diffuse ISM
is generally large enough to prevent the ICM wind from promptly
sweeping it, and leaving bare cores.

It also seems unlikely that the ICM could sweep through many small, as
yet unresolved holes.  Such holes would be connected to their
surroundings by magnetic fields, and as pointed out by Crosthwaithe et
al. they would be closed very quickly by the effects of shear. Indeed,
the shear timescale is comparable to the timescale for the wind to
flush the holes.  Therefore, except for large central holes and gaps
between strong spirals, it seems likely that the disc will respond as
a cohesive whole, whether it is promptly swept, or only compressed.

(4) {\it {Gravitational instability}} is strongly suggested by the
prompt formation of flocculent spirals.  To confirm this, we have
computed the value of the Toomre Q factor in annular rings in the
fiducial model at various times.  We find that $Q \simeq 3$ at most
radii in the initial model, but after the onset of the wind, the Q
value typically drops to less than or about 1.0 at these radii.

Interestingly, the cause of this instability is the displacement of
the disc relative to the halo potential center, which was described in
Section 4.1.  We can understand the triggering of the instability in
terms of the simple gas disc plus halo model used in Section 2.  The
gravitational restoring force balances ram pressure in the midplane of
the displaced disc.  The side of the disc closest to the halo center
of mass (facing the wind) feels an excess ram pressure.  The opposite
side of the disc is farther from the halo center and feels an excess
restoring force.  As a result the disc is squeezed from both sides, in
what amounts to a tidal compression.  These forces are equivalent to
the weight of added mass in the disc, so their effect can be included
in the usual gravitational instability criterion as an 'effective' gas
surface density that is larger than the true surface density.

Assuming that the tidal compression is balanced by a pressure increase
within the disc, we can derive the following expression for the
fractional pressure increase,

\begin{equation}
\frac{{\Delta}P}{P} \simeq 2 
\frac{{\rho_w}{v_w^2}}{{\rho_g}c^2},
\end{equation}
where c is the sound speed of the disc gas, and $\rho_g$ is its mean
density.  The factor of 2 comes from the usual tidal force equation,
and we have also assumed that the disc displacement distance z is
comparable to the disc thickness h.  If $z >> h$, then a factor of h/z
should be inserted in the right hand side of the equation.

If we further assume that the disc gas is isothermal, then the
fractional increase in the effective surface density is also given by
the above expression.  With the following representative values:
${\rho_w}/{\rho_g} = 10^{-4}, v_w \simeq 1000 km/s, c \simeq 5 km/s$,
we get a fractional increase of order unity.  The gas surface density
in most galaxy discs is close to the critical value for gravitational
instability (i.e., Toomre $Q \simeq 1$), so we expect that doubling
the effective surface density will halve Q, and induce a strong
instability, as seen in the simulations.

This instability could not be discovered in cylindrically symmetric
models.  However, it has also not been noted in previous three
dimensional simulations, which did not include radiative cooling.
Heating by the wind would increase the sound speed and the equation
above, reduce disc compression, and prevent gravitational instability.
This prevents annealing (see below) and aids prompt stripping.

 The cooling/heating balance is sensitive to both galactic and ICM
parameters.  We can demonstrate this with a simple estimate.  We take
the kinetic energy flux of the wind ($\rho_{ICM}{v^3}$) as the maximum
heating per unit area.  We equate this to the cooling rate at the peak
of the standard ISM (atomic) cooling curve, multiplied by the disc
thickness (i.e., the cooling per unit surface area of the disc,
$n_{ISM}^2\Lambda_{max}(2h)$).  Using a disc thickness of
$2h \simeq 1.0 kpc.$, the wind density and velocity of our
fiducial model, and $\Lambda_{max} \simeq 5.0 \times 10^{-22} ergs
cm^3 s^{-1}$, we find that cooling dominates whenever $n_{ISM} > 0.02
cm^{-3}$.  Within our fiducial model this critical density is exceeded
by factors of a few to ten in the gas disc.  Thus, energy input from
the wind is never able to heat the disc through the cooling peak.
Moreover, once a bow shock forms, the wind does not directly impact
the disc, and so the heating is likely to be less.  Even stripped
filaments typically exceed the critical density, and they remain at $T
= 10^4 K$ in our models.

Note that the critical density is especially sensitive to the wind
velocity.  E.g., the critical density would
increase to $n_{ISM} \simeq 1.8 cm^{-3}$ with the 'Coma' wind
parameters of Abadi et al., while their 'Virgo' wind is close to our
high wind speed model.  In the case of such a high critical density,
heating would generally dominate, except perhaps, in the densest inner
disc. When heating dominates we would expect that it would facilitate
the stripping of the diffuse ISM, as described below.

(5) When gravitational instability occurs, the spiral waves enhance
the radial tranport of angular momentum in the disc.  The additional
angular momentum deposited at large radii drives expansion, the
resulting column density reduction facilitates stripping.  The loss of
angular momentum at smaller radii leads to compression.  The
'annealed' inner disc within the ring is resistant to stripping.  The
timescale for the combined gravitational instability and annealing is
about $(1-3) \times 10^8 yrs.$, i.e., a few disc dynamical times.

(6) The last stage observable in our models is the continued mass loss
as flocculent spirals in the outer disc shear and expand until their
column density is reduced to the point where they are stripped, as
described in the previous section.  Here we merely emphasize the fact
that since the column density is just low enough for the material to
be lifted, these filaments are very likely to become hung up in the
halo.  Most of this material will eventually be swept, but some falls
back onto the disc.  As shown by Vollmer (2000), this is especially
true when the lingering time is comparable to the cluster crossing
time for galaxies on radial orbits.  In that case, the wind dimishes
before the filaments are swept.

In addition to these sweeping phases, which are clearly discernable in
our simulations, we can speculate about the role of several additional
processes that were not included in our models.  The first
of these is enhanced star formation in the annealed disc.  The
annealed disc has a somewhat higher gas density throughout, but
especially in the ring, so star formation is a natural consequence.
Our models with impulsive heating show that even a burst of star
formation may not lead to much more stripping.  However, the energy
input and subsequent enhanced dissipation may drive some disc
spreading.

Since we find annealing to be most important at ram pressures of order
or somewhat lower than those of the Virgo cluster, annealing induced
star formation may be most important in poor clusters, and young, high
redshift clusters. At the same time, annealing requires some minimum
ram pressure to induce gravitational instability, so annealing induced
star formation is probably less important in galaxy groups than
tidally induced star formation.

One of the best candidates for an annealed disc that we are aware of
is the Virgo spiral NGC 4580.  Its image in the atlas of Sandage \&
Bedke (1994) is remarkably similar to snapshots of the fiducial model
on comparable scales, at late times. Sandage \& Bedke classify it as
type Sc(s)/Sa (!), and state that it is ``peculiar enough to be
outside the classification system.''  It appears to have knots of
young stars thoughout, but concentrated in the broken ring.
Incidently, expanded views of the fiducial model at late times, show
that it's ring is also broken, and is in fact, a spiral that wraps
tightly at that radius. Other examples of possible annealed
morphologies in Virgo can be seen in the paper of Koopmann, Kenney and
Young (2001).

A most impressive set of candidate objects is seen in Figure 11 of
Oemler, Dressler, and Butcher's (1997) study of the four rich clusters
at z=4.  Apparently recent star formation reveals these blue optical
rings.

Star formation and nuclear activity may also be induced by the
fallback of gas clouds stripped from the disc, but not swept out of
the halo. Because the mass involved is generally not large, we would
not expect this process to dominate the global SFR of the galaxy disc,
but the mass could be more than sufficient to feed an AGN. 

\subsection{Annealing Versus Other Instabilities}

As explained in the previous subsection, the annealing process depends
crucially on an induced gravitational instability, and on efficient
angular momentum transport mediated by the spiral waves formed in the
instability.  In this subsection we address two important questions
about these conclusions. The first is, are we sure that this
gravitational instability is the result of ram compression as stated
above? The second is, are there other dynamical instabilities that
might mimick the characteristic gas morphologies formed by
annealing. That is, are those morphologies unique?

In Section 3.2 we noted that a run of the model galaxy with no cluster
wind, confirmed that the model disc remains stable in isolation for
times longer than the characteristic dynamical time (e.g., a free-fall
time). There is no evidence for intrinsic gravitational instability.
There was also no significant change in the global disc angular
momentum in this control run.

This result, together with the consistency of the annealing phenomenon
in the different model runs described in the previous section, and the
analytic theory, provide us a good deal of confidence that we
understand the annealing process.  However, there remains the
possibility that the disc is near the gravitational stability
threshold, so that any significant disturbance, including modest
transients, could trigger the onset of instability.  In this case the
initial wind impact, could be the accidental instability trigger,
rather than the effective weight of steady compression.

We are skeptical of this possibility because we have used this code
with the same initial galaxies in other applications (including models
of galaxy collisions) without triggering internal
instabilities. However, we have also carried out a test simulation
designed to search for internal instability triggered by transient
finite amplitude disturbances.  Specifically, we increased the mass of
all particles by 5\%, and introduced a pseudo-tidal velocity
disturbance, also with a 5\% amplitude.  The latter consisted of an
outward velocity impulse added to the x-component of each velocity
vector, and an inward part added to the y,z velocities, with the
amplitude of each added velocity modulated in latitude as in a pure
tidal perturbation.

\begin{figure}
\includegraphics{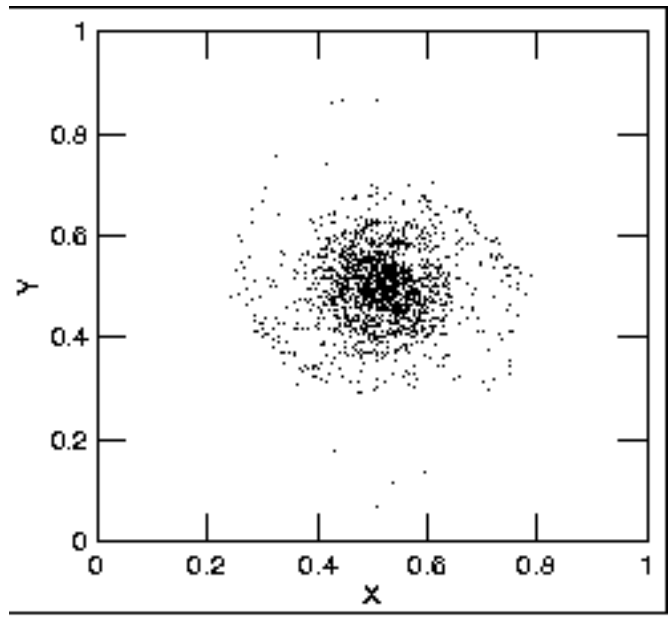}
\caption{Face-on view of a model gas disc at time of 
${1.2 \times 10^8 yrs.}$ after the input of a 
tidal and gravitational disturbance.  This
particular timestep is shown because of its resemblence to annealed
discs.  See text for details.}
\label{spr}
\end{figure}

As expected this disturbance generated spiral and ring waves in the
disc, and these propagating waves persisted through the run.
Generally, these wave morphologies look very different from those of
annealed galaxies.  However, at the timestep show in Figure \ref{spr},
the morphology appears quite similar to those in Figure \ref{c1a}.

In summary, by introducing both a modest tidal and gravitational
perturbation we generated a significant response, and found that it
can mimick the annealing morphology, though only for short times.  The
waves in the test galaxy will eventually decay, leaving a disc that is
little changed from the initial one.  In the annealed galaxies the
changes are permanent.  This is clear from the mass and angular
momentum loss diagrams of the previous section.  In the test
simulation, no mass is lost, and the angular momentum changes by less
than 1\% during the run.

The radial distribution of angular momentum does change slightly in
the test run. Since we have argued that wind compression is like added
weight on the disc, it is likely that by adding enough mass to gas
particles we could replicate the annealing instability without a
wind. However, the estimates of the previous section show that we
would have to increase the disc mass by about 100\%, not the much
smaller amount used in the test.  In any case, it is clear that the
gravitational instability in the annealing models is not the result of
numerical inaccuracy, or inherent instability of the disc to a modest
perturbation.

The test run also suggests that while the morphology of annealed
galaxies is not completely unique, good mimicks are rare.  It is
evidently very difficult to produce this morphology by pure tidal
disturbances. Mass transfer or accretion events may be able to do it,
but again the appearance of such galaxies will generally be dominated
by transients with a very different appearance. 

\subsection{Scaling in Sweeping}

Disc galaxies on radial orbits in clusters will experience most of
their stripping in crossing the cluster core where the ICM densities
and their orbital speeds are highest.  In light of the results above
and previous work, it appears that in a core traversal stripping can
be terminated in one of three ways: 1) in high ram pressure cases most
of the gas is promptly stripped, 2) gravitational instability and
angular momentum transfer cause annealing, 3) rapid core traversal
ends stripping on a timescale shorter than the annealing time.  In
large, dense clusters the annealing and core crossing time are
comparable, but the ram pressure is generally large, so prompt
stripping is dominant. In clusters comparable to Virgo or smaller,
annealing may be important. Annealing will also be important for
galaxies on nearly circular orbits, though in such cases star
formation heating may drive secular mass loss.

\begin{figure}
\includegraphics{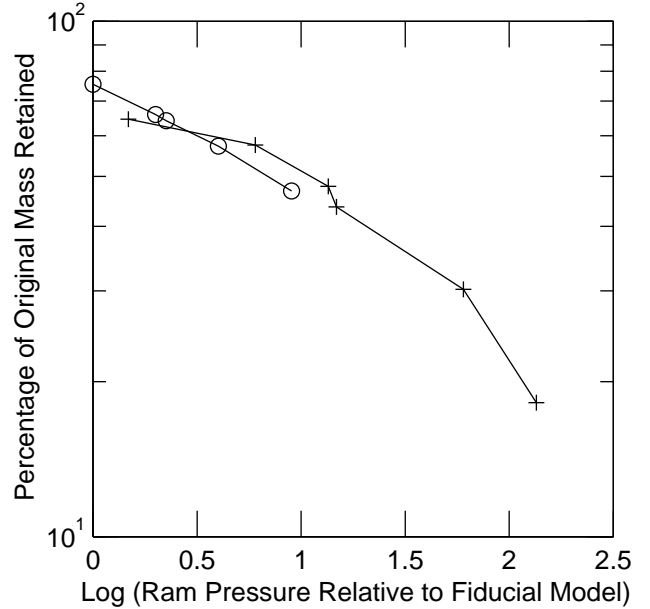}
\caption{Percentage of original gas disc mass retained against face-on
winds, as a function of ram pressure (relative to the fiducial
model). Circles represent the results of our simulations, plus signs
are results from Abadi et al. (1999), normalized as described in the
text.}
\label{rel}
\end{figure}

In most cases, either prompt stripping or annealing dominate, so
stripping goes to saturation and ends on a timescale of order a few
times $10^8$ yrs.  Thus, we can speak of discrete mass loss events,
and ask whether the mass loss in such events is a regular function of
the parameters?  Figure \ref{rel} shows the percentage of gas mass
retained in our simulations (including some additional runs not
described in Section 4) as a function of the ram pressure
${\rho_w}{v_w}^2$ in the wind, normalized to that of the fiducial
model.  This log-log plot shows that the mass retained is in fact a
power-law function of the ram pressure, with an index of about
0.21. The plots includes runs in which the either the wind density and
velocity were changed, but there is no dependence on these parameters
individually, only the ram pressure.

Suppose that the cutoff radius for stripping/sweeping is determined by
a universal critical value of the parameter W(R) of Section 2, i.e.,
$W(R_{cr}) = W_{cr}$.  Since, as noted in Section 2, W is the ratio of
the free-fall time to the momentum change time, this seems to be a
reasonable assumption.  Using the definition of W, we can write,

\begin{equation}
W_{cr} = \frac{\rho_{w}v_{w}^{2}R_{cr}}{{\rho_g}{v_{c}^2}(2h)},
\end{equation}
where ${\rho_g}2h$ is the gas column density written as the
radius-dependent mean disc density times the disc thickness.  In late
type spirals, the gas column density generally falls off as $1/R$, but
we will make the slightly more general assumption that the dependence
is of the form, $R^{-p}$.  We will also assume that the circular
velocity in the disc goes as $R^{-m/2}$, where $m = 0,1$ for flat and
Keplerian rotation curves, respectively.

If we compare identical galaxies in two different cluster
environments, with these assumptions, the ratio of the disc cutoff
radii in the two cases is given by,

\begin{equation}
\frac{R_2}{R_1} = {\biggl(}
\frac{{\rho_{w1}}{v_{w1}^2}}{{\rho_{w2}}{v_{w2}^2}}
{\biggr)}^{1/(1+p+m)}.
\end{equation}

The gas mass out to the critical radius goes as $M_{g}(R) \propto
R^{2-p}$, so that the ratio of retained gas masses is,

\begin{equation}
\frac{M_{g}({R_2})}{M_{g}({R_1})} = 
{\biggl(}\frac{R_2}{R_1}{\biggr)}^{2-p} = 
{\biggl(}
\frac{{\rho_{w1}}{v_{w1}^2}}{{\rho_{w2}}{v_{w2}^2}}
{\biggr)}^{\frac{2-p}{1+p+m}}.
\end{equation} 

Thus, this equation predicts that the index of the mass retention
versus ram pressure power-law is $s = -(2-p)/(1+p+m)$.  For a
flat-rotation curve galaxy, with a gas surface density that falls as
1/R with radius, m=0, p=1, and the index is $s = -0.5$. This is a much
steeper falloff than we find in our simulations, yet this is more or
less what we would expect for prompt stripping from a typical
late-type disc.

In addition, the loss of angular momentum in the annealing process drives
radial compression of the gas disc, and reduces the mass loss.  This
can be viewed as making the effective critical radius in the initial
disc larger.  The loss of angular momentum increases as the ram
pressure increases.  Suppose, for example, that the total angular
momentum loss increases linearly with the ram pressure.  In a m=0, p=1
disc with fixed central surface density, the total angular momentum
scales with the total disc area, or the square of the effective
critical radius.  In this case, the effective critical radius
increases as the square root of ram pressure, and the index s=0, i.e.,
mass loss is independent of ram pressure.

However, the models show a more modest dependence of angular momentum
loss on ram pressure, more like a square root dependence than a
linear.  This leads to $s \simeq 1/4$, which is quite close to the
value $s \simeq 0.21$ derived directly from the models.  Therefore, we
attribute the weak dependence of mass loss on wind parameters to the
annealing process.

In Figure \ref{rel} we have also plotted results from Abadi et
al. (1999), who used ICM densities ranging from $5.64 \times 10^{-24}
g/cm^3$, to an order of magnitude higher, and relative velocities of
$1000-3000 km/s$ (see their Table 2).  The model galaxy used in their
simulations was not identical to ours, so to normalize to our fiducial
model we must multiply by the ratio of the factors
${\rho_g}{v_{c}^2}(2h)$ for the two model galaxies (see eq. (9)).
We have estimated this ratio in a very simple way. First it appears
that the mean gas column densities ${\rho_g}(2h)$ in the two
models are comparable, so we have taken them as equal.  The quantities
${v_{c}^2}$ are assumed to scale with the total mass of the model
galaxies, which have a ratio of about 4.  Serendiptously, this rough
normalization gives almost overlapping curves for the two sets of
models; we might have expected a greater offset. 

The overlap may be due to a partial cancellation of two effects.
Prompt mass loss may be aided by compressional heating in the Abadi et
al. models, and so, is greater. However, gravitational instability and
angular momentum transport drive additional mass loss on longer
timescales in our models.  Apparently, the end result is not greatly
different.  The fact that the form of the Abadi et al. mass retained
versus ram pressure relation is very similar to ours does not depend
on the normalization.  The combined results show that this relation
extends over two orders of magnitude in dimensionless ram pressure.

More generally, the figure indicates that the mass loss from
typical late-type disc galaxies, in a single, saturated, face-on,
stripping event, is determined by the dimensionless parameter,

\begin{equation}
S = {\biggl(}  
\frac{\rho_{w}v_{w}^{2}}{{\rho_d}{v_{c}^2}h}
{\biggr)}{\biggl/}
{\biggl(}
\frac{\rho_{w}v_{w}^{2}}{{\rho_d}{v_{c}^2}h}
{\biggr)}_{fiducial} = 
\frac{\tau_{mom}}{\tau_{mom,fid}}
\end{equation}
i.e., the ratio of the momentum change timescale of Section 2 relative
to that of the fiducial model.

\section{CONCLUSIONS}

In summary, we have presented a small grid of three-dimensional,
self-consistent, thermohydrodynamical models of stripping of gas from
disc galaxies traversing the ICM of a galaxy cluster.  We confirm many
of the general results of other recent simulations.  However, we have
focussed on cooling effects and on models with lower ram pressures
than those presented by Abadi et al. (1999) and Quilis et al. (2000).
As in those works, we maintained a constant ICM wind on the target
disc.  We also ran our simulations for a relatively long time, up to
$6 \times 10^8 yrs.$, primarily to follow dynamical processes to
completion, but also beause this is appropriate for some applications.

We find that stripping is largely deterministic.  If the ram pressure
is great enough to strip most of the disc promptly, then it is also
likely strong enough to sweep the gas directly out of the galaxy halo,
and into the general ICM.  If the ram pressure is not so great, then a
good deal of the gas may become hung-up in the effective potential
minimum described in Section 2, for a time of order $10^8 yrs.,$
before being swept.  A small amount of this material falls back into
the galaxy.  As Vollmer (2000) emphasized, a good deal more would fall
back when the wind duration is short, as in galaxies on radial orbits.

The hang-up effect is one type of longer term mass loss, and may
account for observations of HI clouds and filaments in the wake of
some cluster disc galaxies.  In our models this material is usually
concentrated within a couple galaxy diameters downstream from the
disc.  The location of the effective potential mimimum depends on the
structure of the dark halo of the galaxy.  The dark halos of our model
galaxies were quite concentrated.

Another important process in prolonging the gas removal is the
gravitational instability and inner disc annealing process described
above.  Cooling, and a substantial tidal compression are required for
this instability.  After flocculent spirals form, angular momentum
transport and the shearing of these waves occur on comparable
timescales.  The result is removal of the flocculent spirals from the
outer disc; they blow away as gas filaments.  The inner disc is
radially compressed, and with higher mean column densities it is
'annealed' against further disruption. The boundary of this inner disc
is often marked by a ringlike structure, which may in fact, consist of
tightly wound spirals. The compression may induce subsequent star
formation.

The primary reason these processes were not discovered in previous
simulations is that they are fully three-dimensional, and require
cooling. The role of cooling in retaining gas is somewhat
counter-intuitive.  Stripping criteria do not generally include gas
temperature amoung their various dependencies. However, the role of
cooling is indirect, it allows the onset of gravitational instability,
which promptly generates an annealed or sweeping-resistant inner disc,
as described above.

Despite all the complexities, we find that the net amount of sweeping at
the end of our runs, when the sweeping has indeed terminated, is a
simple power-law function of one dimensionless parameter.  This 'S'
parameter, described in the previous section, includes contributions
from the ICM ram pressure, the galaxy halo (via the maximum circular
velocity), and the gas disc, as might be expected.

These results are modified in cases where ICM wind impacts the gas
disc at a significant angle to the disc symmetry axis.  It is still
true in such cases that there is prompt stripping and downstream
hang-up.  However, in our models we find more material gets hung-up,
and the long term mass loss takes longer than in corresponding face-on
cases.  Of course, the ram pressure is reduced relative to the face-on
case with the same ICM density and relative velocity. However, the
differences are not simply explained by a projection cosine.  This is
illustrated by the coupling between rotation and mass loss discussed
in Section 3.  This coupling is evidently responsible for part of the
time-delay in the mass loss in the angled cases.  In addition, the
loss in angular momentum for a given amount of mass removal was higher
in the tilted models than in face-on models with the same ICM
parameters.  This explains why higher inclination models appear to
have more compact ring structures as well.  Nonetheless, the mass loss
process in the inclined cases is still proceeds in regular stages.

We conclude with a couple more speculations.  The first is that star
formation, and the formation of star clusters or small dwarf galaxies,
may occur in the gas that gets hung-up in the halo, or that is swept
out in relatively dense filaments.  It is tempting to interpret the
dwarf companion (``North Object'') of NGC 1427A, studied by Chaname et
al. 2000, and the dwarf in the wake of NGC 4694 (van Driel \& van
Woerden 1989), as examples.  In the former case, both galaxy and
companion have essentially the same velocity as determined by optical
spectroscopy.  In the latter, there is a peak in the HI wake
superimposed on the optical dwarf.

The fate of star clusters, and dwarf companions formed in this fashion
is not clear.  Small groups or clusters of stars formed in swept
filaments may remain gravitationally attached to the filament and
pulled out of the galaxy, adding to the population that is not bound to
any single galaxy.  Dwarf galaxies formed in the effective potential
minimum, with little angular momentum relative to the parent galaxy,
may separate from their placental gas and fall back into the galaxy.

Our final speculations concern the evolutionary effects of the
stripping power-law derived above.  It is clear from Figure \ref{rel}
that as the ram pressure increases from that of our fiducial model to
that of Abadi et al.'s Coma model, the mass loss in a stripping event
goes from small (25\%) to nearly complete.  If galaxy clusters
complete their growth on a relatively long timescale, via subcluster
mergers, then the ICM density and free-fall velocity increase on a
comparable timescale.  Except in clusters that grow quickly to Coma
proportions, we might expect that sweeping was much less efficient in
clusters at high redshift.  (Note, however, that annealing also
requires a minimum wind pressure to generate the compression needed
for gravitational instability.)

If ICMs do develop on long timescales, galaxies on radial orbits can
experience repeated cycles of sweeping, annealing, and compression
induced starbursts.  This process could contribute to the
Butcher-Oemler effect.  There is evidence that the dominant processes
driving the Butcher-Oemler effect are galaxy collisions and effects
related to the merger of subclusters (e.g., Couch et al. 1994). The
annealing and compression induced starbursts may well be the relevant
process in subcluster mergers.

Successive cycles of annealing may also be an effective means of
building up bulges, and converting spirals to S0s on a timescale of
several cluster crossing times.  The power-law functions for mass and
angular momentum loss in discrete events could be used in simulations
of cluster evolution to investigate these possibilities.

\section{ACKNOWLEDGEMENTS}

We are grateful to D. C. Smith for helpful conversations and computer
tips, to J. van Gorkom and J. Kenney for advice and results in advance
of publication, and to an anonymous referee for many helpful
suggestions.  This research has made use of the NASA/IPAC
Extragalactic Database (NED) which is operated by the Jet Propulsion
Laboratory, California Institute of Technology, under contract with
the National Aeronautics and Space Administration.  This research has
also made use of NASA's Astrophysics Data System Bibliographic
Services.

\end{document}